\documentstyle[12pt]{art12}
\makeatletter
\let\reset@font\empty

\def\indexname{Index}
\def\figurename{Figure}
\def\tablename{Table}
\def\abstractname{Abstract}

\def\@ptsize{0}
\@namedef{ds@11pt}{\def\@ptsize{1}}
\@namedef{ds@12pt}{\def\@ptsize{2}}
\def\ds@twoside{\@twosidetrue
           \@mparswitchtrue}

\def\ds@draft{\overfullrule 5\p@}

\newif\if@titlepage \@titlepagefalse
\def\ds@titlepage{\@titlepagetrue}

\def\ds@twocolumn{\@twocolumntrue}

\newdimen\mathindent
\newif\ifletter
\newif\ifpmb
\newlength{\varind}
\newlength{\figdepth}
\newlength{\figwidth}
\newlength{\secfigwidth}

\newlength{\indentedwidth}

\newcounter{jnl}
\newcounter{yr}
\newcounter{tabtype}
\newcounter{figtype}
\newcounter{eqnval}
\@twosidetrue
\def\ds@draft{\overfullrule 5\p@}
\@options

\def\@normalsize{\@setsize\normalsize{16pt}\xiipt\@xiipt
  \abovedisplayskip 12pt plus3pt minus6pt
  \belowdisplayskip \abovedisplayskip
  \abovedisplayshortskip \z@ plus4pt
  \belowdisplayshortskip 7pt plus4pt minus4pt}
\def\small{\@setsize\small{14pt}\xipt\@xipt
  \abovedisplayskip 10pt plus 3pt minus 4pt
  \belowdisplayskip \abovedisplayskip
  \abovedisplayshortskip \z@ plus3pt
  \belowdisplayshortskip 5pt plus3pt minus 3pt
  \def\@listi{\topsep 5pt plus 3pt minus 3pt\parsep 0pt plus 1pt
         \itemsep \parsep}}
\def\footnotesize{\@setsize\footnotesize{14pt}\xpt\@xpt
  \abovedisplayskip 7pt plus 3pt minus 4pt
  \belowdisplayskip \abovedisplayskip
  \abovedisplayshortskip \z@ plus 2pt
  \belowdisplayshortskip 3pt plus 1pt minus2pt
  \def\@listi{\topsep 4pt plus 2pt minus 2pt\parsep 0pt plus 1pt
         \itemsep \parsep}}
\def\scriptsize{\@setsize\scriptsize{13pt}\ixpt\@ixpt}
\def\tiny{\@setsize\tiny{10pt}\viipt\@viipt}
\def\large{\@setsize\large{18pt}\xivpt\@xivpt}
\def\Large{\@setsize\Large{22pt}\xviipt\@xviipt}
\def\LARGE{\@setsize\LARGE{25pt}\xxpt\@xxpt}
\def\huge{\@setsize\huge{30pt}\xxvpt\@xxvpt}
\def\Huge{\@setsize\Huge{30pt}\xxvpt\@xxvpt}
\normalsize
\skip\footins 12\p@ plus 4\p@ minus 2\p@
\@beginparpenalty -\@lowpenalty
\@endparpenalty -\@lowpenalty
\@itempenalty -\@lowpenalty

\@noskipsecfalse   % new version

\def\section{\@startsection{section}{1}{\z@}{-3.5ex plus -1ex minus
 -.2ex}{2.3ex plus .2ex}{\noindent\reset@font\normalsize\bf\raggedright}}
\def\subsection{\@startsection{subsection}{2}{\z@}{-3.25ex plus -1ex minus
 -.2ex}{1.5ex plus .2ex}{\noindent\reset@font
  \normalsize\it\raggedright\nohyphens}}
\def\subsubsection{\@startsection{subsubsection}{3}{\z@}{-3.25ex plus
-1ex minus -.2ex}{-1em}{\reset@font\normalsize\it\nohyphens}}
\def\paragraph{\@startsection
 {paragraph}{4}{\z@}{3.25ex plus 1ex minus
.2ex}{-1em}{\reset@font\normalsize\it}}
\def\subparagraph{\@startsection
 {subparagraph}{4}{\parindent}{3.25ex plus 1ex minus
 .2ex}{-1em}{\reset@font\normalsize\it}}

\def\@sect#1#2#3#4#5#6[#7]#8{\ifnum #2>\c@secnumdepth
     \let\@svsec\@empty\else
     \refstepcounter{#1}\edef\@svsec{\csname the#1\endcsname.\hskip 1em}\fi
     \@tempskipa #5\relax
      \ifdim \@tempskipa>\z@
        \begingroup #6\relax
          \noindent{\hskip #3\relax\@svsec}{\interlinepenalty \@M #8\par}%
        \endgroup
       \csname #1mark\endcsname{#7}\addcontentsline
         {toc}{#1}{\ifnum #2>\c@secnumdepth \else
                      \protect\numberline{\csname the#1\endcsname}\fi
                    #7}\else
        \def\@svsechd{#6\hskip #3\relax  %% \relax added 2 May 90
                   \@svsec #8\csname #1mark\endcsname
                      {#7}\addcontentsline
                           {toc}{#1}{\ifnum #2>\c@secnumdepth \else
                             \protect\numberline{\csname the#1\endcsname}\fi
                       #7}}\fi
     \@xsect{#5}}
\def\@ssect#1#2#3#4#5{\@tempskipa #3\relax
   \ifdim \@tempskipa>\z@
     \begingroup #4\noindent{\hskip #1}{\interlinepenalty
   \@M #5\par}\endgroup
   \else \def\@svsechd{#4\hskip #1\relax #5}\fi
    \@xsect{#3}}

\def\appendix{\@@par
 \setcounter{section}{0}
 \setcounter{subsection}{0}
 \setcounter{subsubsection}{0}
 \setcounter{equation}{0}
 \setcounter{figure}{0}
 \setcounter{table}{0}
 \def\thesection{Appendix \Alph{section}}
 \def\theequation{\ifnumbysec
      \Alph{section}.\arabic{equation}\else
      \Alph{section}\arabic{equation}\fi}
 \def\thetable{\ifnumbysec
      \Alph{section}\arabic{table}\else
      A\arabic{table}\fi}
 \def\thefigure{\ifnumbysec
      \Alph{section}\arabic{figure}\else
      A\arabic{figure}\fi}}

\leftmargin\leftmargini
\labelwidth\leftmargini\advance\labelwidth-\labelsep
\def\@listI{\leftmargin\leftmargini \parsep 4\p@ plus2\p@ minus\p@
\topsep 8\p@ plus2\p@ minus4\p@
\itemsep 4\p@ plus2\p@ minus\p@}

\let\@listi\@listI
\@listi

\def\@listii{\leftmargin\leftmarginii
 \labelwidth\leftmarginii\advance\labelwidth-\labelsep
 \topsep 3\p@ plus 1\p@ minus 1\p@
 \parsep 0\p@ plus 1\p@
 \itemsep \parsep}
\def\@listiii{\leftmargin\leftmarginiii
 \labelwidth\leftmarginiii\advance\labelwidth-\labelsep
 \topsep 2\p@ plus 1\p@ minus 1\p@
 \parsep \z@ \partopsep 1\p@ plus 0\p@ minus 1\p@
 \itemsep \topsep}
\def\@listiv{\leftmargin\leftmarginiv
 \labelwidth\leftmarginiv\advance\labelwidth-\labelsep}
\def\@listv{\leftmargin\leftmarginv
 \labelwidth\leftmarginv\advance\labelwidth-\labelsep}
\def\@listvi{\leftmargin\leftmarginvi
 \labelwidth\leftmarginvi\advance\labelwidth-\labelsep}

\def\hexnumber@#1{\ifcase#1 0\or 1\or 2\or 3\or 4\or 5\or 6\or 7\or 8\or
 9\or A\or B\or C\or D\or E\or F\fi}
\edef\bffam@{\hexnumber@\bffam}
\mathchardef\bGamma "0\bffam@00
\mathchardef\bDelta "0\bffam@01
\mathchardef\bTheta "0\bffam@02
\mathchardef\bLambda "0\bffam@03
\mathchardef\bXi "0\bffam@04
\mathchardef\bPi "0\bffam@05
\mathchardef\bSigma "0\bffam@06
\mathchardef\bUpsilon "0\bffam@07
\mathchardef\bPhi "0\bffam@08
\mathchardef\bPsi "0\bffam@09
\mathchardef\bOmega "0\bffam@0A

\def\theenumi{\roman{enumi}}

\def\theenumii{\alph{enumii}}
\def\p@enumii{\theenumi.}

\def\theenumiii{\arabic{enumiii}}
\def\p@enumiii{\p@enumii.\theenumii}

\def\p@enumiv{\p@enumiii.\theenumiii}

\def\labelitemi{$\m@th\bullet$}

\def\labelitemiii{$\m@th\ast$}
\def\labelitemiv{$\m@th\cdot$}

\def\verse{\let\\=\@centercr
 \list{}{\itemsep\z@ \itemindent -1.5em\listparindent \itemindent
 \rightmargin\leftmargin\advance\leftmargin 1.5em}\item[]}

\def\quotation{\list{}{\listparindent 1.5em
 \itemindent\listparindent
 \rightmargin\leftmargin\parsep 0\p@ plus 1\p@}\item[]}

\def\descriptionlabel#1{\hspace\labelsep \bf #1}
\def\description{\list{}{\labelwidth\z@ \itemindent-\leftmargin
 \let\makelabel\descriptionlabel}}

\def\enumerate{\ifnum \@enumdepth >3 \@toodeep\else
      \advance\@enumdepth \@ne
      \edef\@enumctr{enum\romannumeral\the\@enumdepth}\list
      {\csname label\@enumctr\endcsname}{\usecounter
        {\@enumctr}\def\makelabel##1{##1\hss}}\fi}
\def\itemize{\ifnum \@itemdepth >3 \@toodeep\else \advance\@itemdepth \@ne
\edef\@itemitem{labelitem\romannumeral\the\@itemdepth}%
\list{\csname\@itemitem\endcsname}{\def\makelabel##1{##1\hss}\topsep=3pt
  \parsep=0pt\listparindent=0pt\itemsep=0pt\partopsep=0pt\rightmargin=0pt
  }\fi}
\tabbingsep \labelsep
\skip\@mpfootins = \skip\footins
\def\titlepage{\@restonecolfalse\if@twocolumn\@restonecoltrue\onecolumn
     \else \newpage \fi \thispagestyle{myheadings}\c@page\z@}

\def\endtitlepage{\if@restonecol\twocolumn \else \newpage \fi}

\newcounter {section}
\newcounter {subsection}[section]
\newcounter {subsubsection}[subsection]
\newcounter {paragraph}[subsubsection]
\newcounter {subparagraph}[paragraph]

\def\thesection {\arabic{section}}

\def\@chapapp{Section}

\def\@pnumwidth{1.55em}
\def\@tocrmarg {2.55em}
\def\@dotsep{4.5}
\def\tableofcontents{\@restonecolfalse\if@twocolumn\@restonecoltrue
 \onecolumn\fi\section*{Contents}{}\thispagestyle{empty}
 \@starttoc{toc}\if@restonecol\twocolumn\fi}
\def\l@section{\@dottedtocline{1}{1.5em}{2.3em}}
\def\l@subsection{\@dottedtocline{2}{3.8em}{3.2em}}
\def\l@subsubsection{\@dottedtocline{3}{7.0em}{4.1em}}
\def\l@paragraph{\@dottedtocline{4}{10em}{5em}}
\def\l@subparagraph{\@dottedtocline{5}{12em}{6em}}
\def\listoffigures{\@restonecolfalse\if@twocolumn\@restonecoltrue\onecolumn
 \fi\section*{List of Figures\@mkboth
 {LIST OF FIGURES}{LIST OF FIGURES}}\@starttoc{lof}\if@restonecol\twocolumn
 \fi}
\def\l@figure{\@dottedtocline{1}{1.5em}{2.3em}}
\def\listoftables{\@restonecolfalse\if@twocolumn\@restonecoltrue\onecolumn
 \fi\section*{List of Tables\@mkboth
 {LIST OF TABLES}{LIST OF TABLES}}\@starttoc{lot}\if@restonecol\twocolumn
 \fi}
\let\l@table\l@figure
\def\@dottedtocline#1#2#3#4#5{\ifnum #1>\c@tocdepth \else
  \vskip \z@ plus .2\p@
  {\leftskip #2\relax \rightskip \@tocrmarg \parfillskip -\rightskip
    \parindent #2\relax\@afterindenttrue
   \interlinepenalty\@M
   \leavevmode
   \@tempdima #3\relax \advance\leftskip \@tempdima
   \hbox{}\hskip -\leftskip
    #4\nobreak\hfill \nobreak \hbox to\@pnumwidth{\hfil
   \rm #5}\@@par}\fi}

\@addtoreset{footnote}{page}
\long\def\@makefntext#1{\parindent 1em\noindent
 \makebox[1em][l]{\footnotesize\rm$\m@th{\fnsymbol{footnote}}$}%
 \footnotesize\rm #1}
\def\@makefnmark{\hbox{${\fnsymbol{footnote}}\m@th$}}
\def\@thefnmark{\fnsymbol{footnote}}
\def\footnote{\@ifnextchar[{\@xfootnote}{\stepcounter{\@mpfn}%
       \begingroup\let\protect\noexpand
       \xdef\@thefnmark{\thempfn}\endgroup
     \@footnotemark\@footnotetext}}
\def\@fnsymbol#1{\ifcase#1\or \dagger\or \ddagger\or \S\or
   \|\or \P\or ^{+}\or ^{\tsty *}\or \sharp
   \or \dagger\dagger \else\@ctrerr\fi\relax}

\def\[{\relax\ifmmode\@badmath\else
 \begin{trivlist}
 \@beginparpenalty\predisplaypenalty
 \@endparpenalty\postdisplaypenalty
 \item[]\leavevmode
 \hbox to\linewidth\bgroup$ \displaystyle
 \hskip\mathindent\bgroup\fi}
\def\]{\relax\ifmmode \egroup $\hfil \egroup \end{trivlist}\else \@badmath \fi}
\def\equation{\@beginparpenalty\predisplaypenalty
 \@endparpenalty\postdisplaypenalty
\refstepcounter{equation}\trivlist \item[]\leavevmode
 \hbox to\linewidth\bgroup $ \displaystyle
\hskip\mathindent}
\def\endequation{$\hfil \displaywidth\linewidth\@eqnnum\egroup \endtrivlist}
\def\eqnarray{\stepcounter{equation}\let\@currentlabel=\theequation
\global\@eqnswtrue
\global\@eqcnt\z@\tabskip\mathindent\let\\=\@eqncr
\abovedisplayskip\topsep\ifvmode\advance\abovedisplayskip\partopsep\fi
\belowdisplayskip\abovedisplayskip
\belowdisplayshortskip\abovedisplayskip
\abovedisplayshortskip\abovedisplayskip
$$\halign to
\linewidth\bgroup\@eqnsel$\displaystyle\tabskip\z@
 {##{}}$&\global\@eqcnt\@ne $\displaystyle{{}##{}}$\hfil    %\hfil delete
 &\global\@eqcnt\tw@ $\displaystyle{{}##}$\hfil
 \tabskip\@centering&\llap{##}\tabskip\z@\cr}
\def\endeqnarray{\@@eqncr\egroup
 \global\advance\c@equation\m@ne$$\global\@ignoretrue }
\newcommand{\jl}[1]{\setcounter{jnl}{#1}%
    \ifnum\thejnl=12\global\pmbtrue\fi
    \ifnum\thejnl=15\global\pmbtrue\fi}
\def\journal{\ifnum\thejnl=1 J. Phys.\ A: Math.\ Gen.\
        \else\ifnum\thejnl=2 J. Phys.\ B: At.\ Mol.\ Opt.\ Phys.\
        \else\ifnum\thejnl=3 J. Phys.:\ Condens. Matter\
        \else\ifnum\thejnl=4 J. Phys.\ G: Nucl.\ Part.\ Phys.\
        \else\ifnum\thejnl=5 Inverse Problems\
        \else\ifnum\thejnl=6 Class. Quantum Grav.\
        \else\ifnum\thejnl=7 Network\
        \else\ifnum\thejnl=8 Nonlinearity\
        \else\ifnum\thejnl=9 Quantum Opt.\
        \else\ifnum\thejnl=10 Waves in Random Media\
        \else\ifnum\thejnl=11 Pure Appl. Opt.\
        \else\ifnum\thejnl=12 Phys. Med. Biol.\ %
        \else\ifnum\thejnl=13 Modelling Simul.\ Mater.\ Sci.\ Eng.\
        \else\ifnum\thejnl=14 Plasma Phys. Control. Fusion\
        \else\ifnum\thejnl=15 Physiol. Meas.\
        \else\ifnum\thejnl=16 Sov.\ Lightwave Commun.\
        \else\ifnum\thejnl=17 High Perform.\ Polym.\
        \else\ifnum\thejnl=18 J.\ Hard Mater.\
        \else\ifnum\thejnl=19 J.\ Phys.\ D: Appl.\ Phys.\
        \else\ifnum\thejnl=20 Supercond.\ Sci.\ Technol.\
        \else\ifnum\thejnl=21 Semicond.\ Sci.\ Technol.\
        \else\ifnum\thejnl=22 Nanotechnology\
        \else\ifnum\thejnl=23 Meas.\ Sci.\ Technol.\
        \else\ifnum\thejnl=24 Plasma Source Sci.\ Technol.\
        \else\ifnum\thejnl=25 Smart Mater.\ Struct.\
        \else\ifnum\thejnl=26 J.\ Micromech.\ Microeng.\
        \else\ifnum\thejnl=27 Distrib.\ Syst.\ Engng\
\else Institute of Physics Publishing
\fi\fi\fi\fi\fi\fi\fi\fi\fi\fi\fi\fi\fi\fi\fi
\fi\fi\fi\fi\fi\fi\fi\fi\fi\fi\fi\fi}

\def\catchline{\hfill}

\def\cpyrtline{\hfill}
\def\maketitle{\vspace*{\baselineskip}\vspace{0\p@ plus1fil}
    \noindent Short title: \@shorttitle\par
    \@submitted
    \vspace*{\baselineskip}
    \noindent\today\par\newpage}
\def\@rticle#1#2{\thispagestyle{myheadings}%
     \vspace*{.5pc}%
    {\parindent=\mathindent \bf #1\par}%
     \vspace*{1.5pc}%
    {\exhyphenpenalty=10000\hyphenpenalty=10000
     \Large\raggedright\noindent
     \bf#2\par}\def\@shorttitle{#1}\futurelet\next\sh@rttitle}%
\def\title#1{\def\@shorttitle{#1}%
    \thispagestyle{myheadings}%
    \vspace*{3pc}{\exhyphenpenalty=10000\hyphenpenalty=10000
    \Large\raggedright\noindent
    \bf#1\par}\futurelet\next\sh@rttitle}

\def\article#1#2{\@rticle{#1}{#2}}
\def\review#1{\@rticle{REVIEW \ifpmb\else ARTICLE\fi}{#1}}
\def\topical#1{\@rticle{TOPICAL REVIEW}{#1}}
\def\ireview#1{\@rticle{INTRODUCTORY REVIEW}{#1}}
\def\comment#1{\@rticle{COMMENT}{#1}}
\def\note#1{\@rticle{NOTE}{#1}}
\def\prelim#1{\@rticle{PRELIMINARY COMMUNICATION}{#1}}
\def\letter#1{\@rticle{LETTER TO THE EDITOR}{#1}}
\def\sh@rttitle{\ifx\next[\let\next=\sh@rt
                \else\let\next=\f@ll\fi\next}
\def\sh@rt[#1]{\gdef\@shorttitle{#1}}
\def\f@ll{}
\renewcommand{\author}[1]{\vspace*{1.5pc}%
   \begin{indented}%
   \item[]\normalsize\ifnum\thejnl=8\bf\else\rm\fi\raggedright#1
   \end{indented}%
   \smallskip}
\def\abstract{\vspace{16pt plus3pt minus3pt}
   \begin{indented}
   \item[]{\bf \abstractname.}\quad\rm\ignorespaces}%
\def\endabstract{\end{indented}\vspace{18\p@ plus18\p@}}
\def\submitted{\def\@submitted{\vspace{\baselineskip}%
     \noindent Submitted to: \journal\par}}
\def\@submitted{}%    Default value
\def\nosections{\vspace{30\p@ plus12\p@ minus12\p@}
    \noindent\ignorespaces}
\def\ack{\ifletter\bigskip\noindent\ignorespaces\else
    \section*{Acknowledgments}\fi}

\newif\ifnumbysec
\def\theequation{\ifnumbysec
      \arabic{section}.\arabic{equation}\else
      \arabic{equation}\fi}
\def\eqnobysec{\numbysectrue\@addtoreset{equation}{section}}
\newenvironment{ceqno}{\begin{ceqno}}{\end{ceqno}}
\def\ceqno{\begin{equation}\begin{array}{@{}*{4}{l}}\dsty}
\def\endceqno{\end{array}\end{equation}}
\def\eqalign#1{\null\vcenter{\def\\{\cr}\openup\jot\m@th
  \ialign{\strut$\displaystyle{##}$\hfil&$\displaystyle{{}##}$\hfil
      \crcr#1\crcr}}\,}
\def\eqalignno#1{\displ@y \tabskip\z@skip
  \halign to\displaywidth{\hspace{5pc}$\@lign\displaystyle{##}$%
    \tabskip\z@skip
    &$\@lign\displaystyle{{}##}$\hfill\tabskip\@centering
    &\llap{$\@lign\hbox{\rm##}$}\tabskip\z@skip\crcr
    #1\crcr}}
\def\cases#1{%
     \left\{\,\vcenter{\def\\{\cr}\normalbaselines\openup1\jot\m@th%
     \ialign{\strut$\displaystyle{##}\hfil$&\tqs
     \rm##\hfil\crcr#1\crcr}}\right.}%
\def\tabular{\def\@halignto{}\@tabular}
\newcommand{\Table}[1]{\def\t@blecap{\caption{#1}}%
   \setcounter{tabtype}{1}\futurelet\next\t@bplace}
\newcommand{\widetable}[1]{\def\t@blecap{\caption{#1}}%
   \setcounter{tabtype}{2}\futurelet\next\t@bplace}
\newcommand{\fulltable}[1]{\def\t@blecap{\caption{#1}}%
   \setcounter{tabtype}{3}\futurelet\next\t@bplace}%
\def\t@bplace{\ifx\next[\let\next=\@tabpl
                 \else\let\next=\@tabnopl\fi\next}
\def\@tabpl[#1]{\begin{table}[#1]\@t@bsize}
\def\@tabnopl{\begin{table}\@t@bsize}
\def\@t@bsize{\ifnum\thetabtype=3\begin{varindent}{0pt}%
   \else\begin{varindent}{\mathindent}\fi
   \t@blecap\lineup\item[]
   \ifnum\thetabtype=1
        \begin{tabular}{@{}l*{15}{l}}
   \else\ifnum\thetabtype=2
        \begin{tabular*}{\indentedwidth}{@{}l*{15}{@{\extracolsep{0pt
plus12pt}}l}}
   \else\begin{tabular*}{\textwidth}{@{}l*{15}{@{\extracolsep{0pt plus12pt}}l}}
   \fi\fi}
\def\endtab{\ifnum\thetabtype=1\end{tabular}
   \else\end{tabular*}\fi\end{varindent}\end{table}}

\def\lineup{\def\0{\hbox{\phantom{\footnotesize\rm 0}}}%
    \def\m{\hbox{$\phantom{-}$}}%
    \def\-{\llap{$-$}}}
\long\def\@makecaption#1#2{\vskip 10\p@
 \ifnum\thefigtype=2\begin{varindent}{\@figindent}
 \item[]{\bf #1.} #2
 \end{varindent}\else
 \ifnum\thefigtype=3
 \footnotesize\rm{\bf #1.} #2\else
 \begin{indented}
 \item[]{\bf #1.} #2
 \end{indented}\fi\fi}
\newcommand{\Figure}[1]{\setcounter{figtype}{1}%
    \def\figspace{}\def\figcap{\caption{#1}}%
    \futurelet\next\@figplace}
\def\@figplace{\ifx\next[\let\next=\@figpl
                 \else\let\next=\@fignopl\fi\next}
\def\@figpl[#1]{\begin{figure}[#1]
   \figspace
   \figcap
   \end{figure}}
\def\@fignopl{\begin{figure}
   \figspace
   \figcap
   \end{figure}}

\newcommand{\sidecap}[3]{\setcounter{figtype}{2}%
    \setlength{\figdepth}{#1}\def\@figindent{#2}%
    \def\sidedc@p{\caption{#3}}%
    \futurelet\next\@sidecapplace}
\def\@sidecapplace{\ifx\next[\let\next=\@sidecappl
                 \else\let\next=\@sidecapnopl\fi\next}
\def\@sidecappl[#1]{\begin{figure}[#1]
    \vbox to\figdepth{\vfill
    \sidedc@p}%
    \setcounter{figtype}{1}\end{figure}}
\def\@sidecapnopl{\begin{figure}
    \vbox to\figdepth{\vfill
    \sidedc@p}%
    \setcounter{figtype}{1}\end{figure}}
\newcommand{\side}[3]{\setcounter{figtype}{3}%
    \setlength{\figdepth}{#1}\setlength{\figwidth}{15pc}
    \setlength{\secfigwidth}{15pc}
    \def\firstc@p{\caption{#2}}\def\secondc@p{\caption{#3}}
    \futurelet\next\@sideplace}
\def\@sideplace{\ifx\next[\let\next=\@sidepl
                 \else\let\next=\@sidenopl\fi\next}
\def\@sidepl[#1]{\begin{figure}[#1]
    \vspace*{1.5pc}\vspace*{\figdepth}
    \parbox[t]{\figwidth}{\firstc@p}\hspace*{1pc}%
    \parbox[t]{\secfigwidth}{\secondc@p}
    \setcounter{figtype}{1}\end{figure}}
\def\@sidenopl{\begin{figure}
    \vspace*{1.5pc}\vspace*{\figdepth}
    \parbox[t]{\figwidth}{\firstc@p}\hspace*{1pc}%
    \parbox[t]{\secfigwidth}{\secondc@p}
    \setcounter{figtype}{1}\end{figure}}
\newcommand{\varside}[4]{\setcounter{figtype}{3}%
    \setlength{\figdepth}{#1}\setlength{\figwidth}{#2}%
    \setlength{\secfigwidth}{30pc}
    \addtolength{\secfigwidth}{-\figwidth}
    \def\firstc@p{\caption{#3}}\def\secondc@p{\caption{#4}}
    \futurelet\next\@sideplace}

\newcounter{figure}
\def\thefigure{\@arabic\c@figure}
\def\fps@figure{htbp}
\def\ftype@figure{1}
\def\ext@figure{lof}
\def\fnum@figure{\figurename~\thefigure}
\def\figure{\@float{figure}}
\let\endfigure\end@float
\@namedef{figure*}{\@dblfloat{figure}}
\@namedef{endfigure*}{\end@dblfloat}
\newcounter{table}
\def\thetable{\@arabic\c@table}
\def\fps@table{htbp}
\def\ftype@table{2}
\def\ext@table{lot}
\def\fnum@table{\tablename~\thetable}
\def\table{\@float{table}}
\let\endtable\end@float
\@namedef{table*}{\@dblfloat{table}}
\@namedef{endtable*}{\end@dblfloat}

\def\thebibliography#1{\list
 {\hfil[\arabic{enumi}]}{\topsep=0\p@\parsep=0\p@
 \partopsep=0\p@\itemsep=0\p@
 \labelsep=5\p@\itemindent=-10\p@
 \settowidth\labelwidth{\footnotesize[#1]}%
 \leftmargin\labelwidth
 \advance\leftmargin\labelsep
 \advance\leftmargin -\itemindent
 \usecounter{enumi}}\footnotesize
 \def\newblock{\ }
 \sloppy\clubpenalty4000\widowpenalty4000
 \sfcode`\.=1000\relax}

\def\numrefs#1{\begin{thebibliography}{#1}}
\def\endnumrefs{\end{thebibliography}}

\def\thereferences{\list{}{\topsep=0\p@\parsep=0\p@
 \partopsep=0\p@\itemsep=0\p@\labelsep=0\p@\itemindent=-18\p@
\labelwidth=0\p@\leftmargin=18\p@
}\footnotesize\rm
\def\newblock{\ }
\sloppy\clubpenalty4000\widowpenalty4000
\sfcode`\.=1000\relax
}
\newenvironment{harvard}{\list{}{\topsep=0\p@\parsep=0\p@
\partopsep=0\p@\itemsep=0\p@\labelsep=0\p@\itemindent=-18\p@
\labelwidth=0\p@\leftmargin=18\p@
}\footnotesize\rm
\def\newblock{\ }
\sloppy\clubpenalty4000\widowpenalty4000
\sfcode`\.=1000\relax}{\endlist}
\def\refs{\begin{harvard}}
\def\endrefs{\end{harvard}}
\newenvironment{indented}{\begin{indented}}{\end{indented}}
\newenvironment{varindent}[1]{\begin{varindent}{#1}}{\end{varindent}}
\def\indented{\list{}{\itemsep=0\p@\labelsep=0\p@\itemindent=0\p@
   \labelwidth=0\p@\leftmargin=\mathindent\topsep=0\p@\partopsep=0\p@
   \parsep=0\p@\listparindent=15\p@}\footnotesize\rm}

\def\varindent#1{\setlength{\varind}{#1}%
   \list{}{\itemsep=0\p@\labelsep=0\p@\itemindent=0\p@
   \labelwidth=0\p@\leftmargin=\varind\topsep=0\p@\partopsep=0\p@
   \parsep=0\p@\listparindent=15\p@}\footnotesize\rm}

\def\tabnotes{\ifnum\thetabtype=1\end{tabular}\else\end{tabular*}\fi}
\def\endtabnotes{\end{varindent}\end{table}}

\newif\if@restonecol
\def\theindex{\@restonecoltrue\if@twocolumn\@restonecolfalse\fi
\columnseprule \z@
\columnsep 35\p@\twocolumn[\section*{\indexname}]%
    \@mkboth{{\indexname}}{{\indexname}}%
    \parindent\z@
    \parskip\z@ plus.3\p@\relax\let\item\@idxitem}
\def\@idxitem{\par\hangindent 30\p@}
\def\subitem{\par\hangindent 30\p@ \hspace*{10\p@}}
\def\subsubitem{\par\hangindent 30\p@ \hspace*{20\p@}}
\def\endtheindex{\if@restonecol\onecolumn\else\clearpage\fi}
\def\indexspace{\par \vskip 10\p@ plus 5\p@ minus 3\p@\relax}

\mark{{}{}}

\def\ps@headings{\let\@mkboth\markboth
 \def\@oddfoot{}%
 \def\@evenfoot{}%
 \def\@evenhead{\makebox[\mathindent][l]{\normalsize\rm \thepage}%
  \normalsize\it\rightmark\hfill}%
 \def\@oddhead{\makebox[\mathindent][r]{\hfill}{\normalsize\it\leftmark}\hfill
  \normalsize\rm\thepage}%
}%

\def\ps@myheadings{\let\@mkboth\markboth
 \def\@oddhead{\catchline}%
 \def\@oddfoot{\cpyrtline}%
 \def\@evenhead{}%
 \def\@evenfoot{}%
}

\def\today{\ifcase\month\or
 January\or February\or March\or April\or May\or June\or
 July\or August\or September\or October\or November\or December\fi
 \space\number\day, \number\year}

\def\@begintheorem#1#2{\rm \trivlist \item[\hskip \labelsep{\it #1\ #2.}]}
\def\@opargbegintheorem#1#2#3{\rm \trivlist
      \item[\hskip \labelsep{\it #1\ #2\ (#3).}]}

\def\p@LaTeX{{L\kern-.3em\lower.1em\hbox{$^{\rm A}$}\kern-.15em%
    T\kern-.1667em\lower.7ex\hbox{E}\kern-.125emX}}
\newcommand{\text}[1]{\mbox{#1}}

\newcommand{\nohyphens}{\hyphenpenalty=10000\exhyphenpenalty=10000}

\renewcommand{\d}{{\mathrm d}}
\renewcommand{\qquad}{\hspace*{25pt}}
\newcommand{\tqs}{\hspace*{25pt}}
\newcommand{\fl}{\hspace*{-\mathindent}}

\def\pt(#1){({\it #1\/})}
\newcommand{\dsty}{\displaystyle}
\newcommand{\tsty}{\textstyle}

\def\;{\protect\psemicolon}
\def\psemicolon{\relax\ifmmode\mskip\thickmuskip\else\kern .3333em\fi}

\newcommand{\opencirc}{\raisebox{2\p@}{\;\circle{5}}}

\newcommand{\fullcirc}{\raisebox{-2\p@}{\Large$\bullet$}}

\newcommand{\boldarrayrulewidth}{1\p@}
\def\bhline{\noalign{\ifnum0=`}\fi\hrule \@height
\boldarrayrulewidth \futurelet \@tempa\@xhline}

\def\@xhline{\ifx\@tempa\hline\vskip \doublerulesep\fi
      \ifnum0=`{\fi}}

\newcommand{\br}{\ms\bhline\ms}
\newcommand{\mr}{\ms\hline\ms}
\newcommand{\ms}{\noalign{\vspace{3\p@ plus2\p@ minus1\p@}}}
\newcommand{\bs}{\noalign{\vspace{6\p@ plus2\p@ minus2\p@}}}
\newcommand{\ns}{\noalign{\vspace{-3\p@ plus-1\p@ minus-1\p@}}}
\newcommand{\es}{\noalign{\vspace{6\p@ plus2\p@ minus2\p@}}\displaystyle}
\newcommand{\JPA}{{\em J. Phys. A: Math. Gen.} }
\newcommand{\PRL}{{\em Phys. Rev. Lett.} }

\ps@headings  \onecolumn
\makeatother
\begin{document}
\jl{1}
\input{epsf.sty}
\eqnobysec
%%%%%%%%%%%%%%%%%%%%%%%%%%%%%%%%%%%%%%%%%%%%%%%%%%%%%%%%%%
\def\hc{h_{c}}
\def\H{H}
\def\Ts{T_{\rm short}}
\def\Tl{T_{\rm long}}
\def\hp{\frac{1+h}{2}}
\def\hm{\frac{1-h}{2}}
\def\M{d}
\def\de{d}
\def\d{{\rm d\/}}
\def\U{U}
\def\F{F}
\def\Ps{|P_0\!>}
\def\FEF{FEF}
%%%%%%%%%%%%%%%%%%%%%%%%%%%%%%%%%%%%%%%%%%%%%%%%%%%%%%%%%%
%%%%%%%%%%%%%%%%%%%%%%%%%%%%%%%%%%%%%%%%%%%%%%%%%%%%%%%%%%
\vspace*{0.5cm}
\begin{center}
   {\bf \Large 
Metastability and spinodal points\\
for a random walker on a triangle
 	}
   \\[15mm]
Peter F. Arndt and
Thomas Heinzel
\\[7mm]
Physikalisches Institut, Nu{\ss}allee 12,
53115 Bonn, Germany\\[3mm]
{peter@theoa1.physik.uni-bonn.de}\\
{tom@maple.physik.uni-bonn.de}
\\[1.5cm]
%{\bf Abstract}
\end{center}
\renewcommand{\thefootnote}{\arabic{footnote}}
\addtocounter{footnote}{-1}
\vspace*{2mm}
%
% Abstract
%
We investigate time-dependent properties of
a single particle model in which a random walker moves on
a triangle and is subjected to non-local
boundary conditions. 
This model exhibits spontaneous breaking of a {\boldmath $Z_2$}
symmetry. The reduced size of the configuration space (compared to related 
many-particle models that also show spontaneous symmetry breaking)
allows us to study the spectrum of the time-evolution operator.
We break the symmetry explicitly and find a stable phase, and a meta-stable
phase which vanishes at a spinodal point.
At this point, the spectrum of the time evolution operator
has a gapless and universal band of excitations with 
a dynamical critical exponent
$z=1$. Surprisingly,
the imaginary parts of the eigenvalues $E_j(L)$ are equally spaced, 
following the rule ${\rm Im}E_j(L)\propto j/L$.
Away from the spinodal point,
we find two time scales in the spectrum.
These results are related to scaling functions for the mean path of the random walker 
and to first passage times.
For the spinodal point, we find
universal scaling behavior.
A simplified version of the
model which can be handled analytically is also presented.
\vspace*{1.5cm}
\begin{flushleft}
cond-mat/9710287
\\
October 1997\\[1cm]
\end{flushleft}
\thispagestyle{empty}
\mbox{}
\newpage
\setcounter{page}{1}
%%%%%%%%%%%%%%%%%%%%%%%%%%%%%%%%%%%%%%%%%%%%%%%%%%%%%%%%%%
\section{Introduction}
Spontaneous symmetry breaking in non-equilibrium
statistical mechanics was recently observed in several
one-dimensional many-particle models \cite{orig,DeDoMu,ScDo,DeEvHaPa,AHR}.
In this paper we consider a single particle random walker model \cite{toy}
which exhibits symmetry breaking of a ${\bf Z}_2$ symmetry.
We investigate the spectrum of the time evolution operator,
and stationary and time-dependent properties.
The random walker moves in the two-dimensional geometry of
a discretized right-angled triangle.
In the interior of the triangle it may hop locally to neighbouring sites,
whereas on the two short sides it may
jump non-locally to a corner.
Unusually for a one particle model the simple model, studied here 
shows spontaneous
symmetry breaking.
Considering the infinite volume limit, the particle
stays in one corner of the triangle and no longer reaches 
the opposite corner which is reflected under the symmetry.
This yields two equally stable phases.

Breaking the symmetry explicitly in a soft way, one of the 
stable phases first becomes meta-stable, before 
one reaches a spinodal point where it becomes unstable.
Hence one can investigate metastability and spinodal points
in non-equilibrium processes with the simple random walker model.

The random walker on a triangle (RWT) model is defined as follows.
The walker moves on the triangle
\begin{eqnarray*} 
{\cal T}_L=\{(j,k)\mid j,k \mbox{ integers}\,>0\mbox{ with }j+k\leq L\}\;.
\end{eqnarray*}
We define
the following processes for the random walker at site $(j,k)$
(see also Ref.\cite{toy}),
where $p$ is the probability of a transition
in an infinitesimal time step $\d t$:
\begin{eqnarray}
(j,k)\rightarrow (j-1,k)\qquad &\mbox{if $j>1$ with }p=a(1+h)\d t
\nonumber\\
(j,k)\rightarrow (j,k-1)\qquad &\mbox{if $k>1$ with }p=a(1-h)\d t
\nonumber\\
(j,k)\rightarrow (j-1,k+1)\qquad &\mbox{if $j>1$ with }p=b(1+h)\d t
\nonumber\\
(j,k)\rightarrow (j+1,k-1)\qquad &\mbox{if $k>1$ with }p=b(1-h)\d t
\nonumber\\
(1,k)\rightarrow (1,L-1)\qquad &\mbox{with }p=c (1+h)\d t
\nonumber\\
(j,1)\rightarrow (L-1,1)\qquad &\mbox{with }p=c (1-h)\d t \;\;.
\label{toy1}
\end{eqnarray}
The first four processes are local hops to neighbouring sites on the triangle,
the last two are non-local jumps to the corners $(1,L-1)$ and $(L-1,1)$ 
respectively. 
For $h=0$ the definitions are invariant under
reflection on the line $j=k$.
We fix the unit of time choosing $c=1/2$. This renders the
probability rates dimensionless.

In the following we also impose the condition $a+b=1/2$, 
which makes the RWT model the 
``zero-temperature'' limit
of a three-state model \cite{orig,toy} which shows spontaneous $CP$ symmetry
breaking.
(Therefore the reflection symmetry of the RWT model will also be denoted as $CP$ symmetry in the following.)
\setlength{\unitlength}{10pt}
\def\shoi{\begin{picture}(25,10)(0,0)
	\thicklines
	\put(0,5){\vector(1,0){20}}
	\put(0,0){\vector(0,1){10}}
	\put(19,4){\makebox(0,0)[c]{$\scriptstyle T$}}
	\put(12,4){\makebox(0,0)[c]{$\scriptstyle T_c$}}
	\put(-1,9.5){\makebox(0,0)[c]{$\scriptstyle h$}}
	\put(-1,8.0){\makebox(0,0)[c]{$\scriptstyle h_c$}}
	\thinlines
        \put(0,8){\line(4,-1){12}}
        \put(0,2){\line(4,1){12}}
	\put(3,4){\makebox(0,0)[c]{A}}
	\put(15,7){\makebox(0,0)[c]{B}}
	\end{picture}
	}
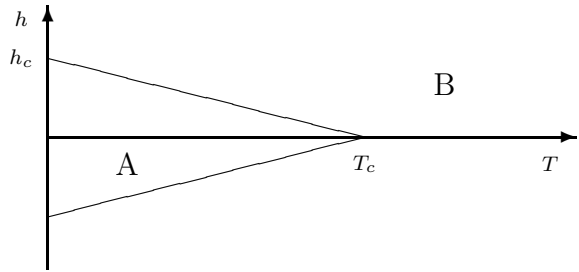
\begin{figure}[tb]
\setlength{\unitlength}{10pt}
\begin{picture}(26,12)(0,0)
\put(10,-1){
        \makebox{\shoi}
        }
\end{picture}
\caption{
\label{figphase}
Simplified phase diagram of the three-state model. 
In the ``low-temperature'' regime A the 
symmetry is spontaneously broken for $h=0$ 
where $h$ is the strength of a symmetry breaking field.
It is separated 
by a line of spinodal points from the ``high-temperature'' regime B.
}
\end{figure}
It would be interesting to study the general case $a+b\neq 1/2$ as well.
In the three-state model, positive and negative particles hop
among vacancies on a one-dimensional chain.
The $C$ operation interchanges positive and negative particles,
and the $P$ operation interchanges left and right on the chain.
The investigation of the free energy functional, and 
of the flip times between stable and meta-stable phases, leads to the
phase structure given in figure \ref{figphase} \cite{us}.
In the $CP$-symmetric case, one finds two broken phases \cite{BeGr} for 
``low temperatures'' (in this model the output 
rates of the particles play the role of a temperature) 
and a disordered phase for ``temperatures'' above $T_c$.
At the critical point ($T=T_c,h=0$), one finds a dynamical
critical exponent $z=2$, which is connected with the appearance
of shocks \cite{us,JaLe,cxxvi,cxxvii}.
Breaking the symmetry explicitly, one finds
a ``low-temperature'' regime A with a stable and a metastable phase.
This regime is separated by
a spinodal line from the regime B (Fig.\ref{figphase}).
At the spinodal line, a previous investigation suggested $z=1$ for $h\neq 0$ \cite{us}.
This motivated us to study time-dependent properties, metastability and 
spinodal points in more detail.
All these are manifestly associated with
non-equilibrium models \cite{PrSc}.

Another model which shows spontaneous $CP$ symmetry breaking
is the two-state (particles and vacancies) 
asymmetric exclusion model \cite{DeDoMu,ScDo,DeEvHaPa}.
There, the $C$ operation interchanges particles and vacancies.
In the spontaneously broken stationary state,
shock profiles (connected with a dynamical critical exponent
$z=2$) appear.
But breaking the symmetry in a soft way one does not find 
a region with a meta-stable phase, and hence there are no
spinodal points present in the model.

%As mentioned above, the RWT model with the above conditions
%on the rates is the zero temperature limit of the three-state model.
%Hence, a spinodal point is also present in the phase diagram of the 
%random walker model.
The time evolution of the RWT model is given by the master equation \cite{EsRi}
\begin{equation}
\label{maeq}
\label{maeqmatrix}
\frac{\d}{\d t}|P\!>=-H\,|P\!>\;\;,
\end{equation}
where $|P\!>$ is the vector of the probabilities $P(j,k)$ of finding the 
random walker at site $(j,k)$.
The ``hamiltonian'' $H$ is the time evolution operator and is determined by 
the processes (\ref{toy1}).
The stationary properties of the system are given by the 
ground state $|P_0\!\!>$ with $H\,|P_0\!\!>\,=0$, and the dynamical properties are given
by the 
excitations of $H$.

The configuration space of many-particle models grows exponentially with the system size
in contrast to
the configuration space of the RWT model which has dimension $L(L-1)/2$.
This polynomial dependence on $L$ 
enabled us to investigate the spectrum
of the model.
By numerical diagonalization of the hamiltonian,
we found an interesting two band structure of the spectrum,
which leads to two gaps or time-scales.
At the spinodal point $h_c=a/(1-a)$, one scale vanishes 
with critical exponent $x=2$ and the 
lower part of the spectrum has the form $E_j=a k_j/L$.
The constants $k_j$ are universal, and do not depend on the remaining 
parameter $a$.
This reminds one of the critical spectrum of a quantum spin chain
related to a conformal field theory.
As a surprise, the imaginary parts of the constants $k_j$ have the form
${\rm Im} k_j=j {\rm Im} k_1$. For the real parts, we do not find any regularity.

The value of $h_c$ has already been conjectured in \cite{toy}.
It is also predicted by taking
the ``zero-temperature'' limit of the mean-field results
for the three-state model discussed in \cite{us}. There, mean-field theory
was found to be compatible with simulations for small ``temperatures''
(output rates).

Besides the study of the time evolution operator one can,
as with other models, investigate the free energy functional
(FEF) of the RWT model.
This functional $f$ is defined as \cite{us,freee}:
\begin{equation}
\label{deffreeenergy}
f_L(\M)=-\frac1L\log P_0(\M,L)\; , \qquad 
f(\M)=\lim_{L\rightarrow\infty}f_L(\M)\; ,
\end{equation}
where $\M$ is an order parameter of the model
and $P_0(\M,L)$ denotes the corresponding stationary probability distribution \cite{us}.
For the RWT model, we take
\begin{equation}
\label{orpa}
\M=(j-k)/L
\end{equation}
as the order parameter.
 
Analogously to the ``low-temperature'' regime of the three-state model,
one finds that the FEF of the RWT model has two global minima ($d=-1,+1$) in the
$CP$-symmetric case ($h=0$).
In the limit $L\rightarrow\infty$, the symmetry is spontaneously
broken and there are two equally stable phases with $d=\pm 1$.
For $0<h<h_c$, the dynamics (\ref{toy1}) favors the movement to
lower values of the order parameter and
the FEF has a global minimum ($d=-1$) and a local minimum ($d=+1$).
This leads to the existence of a stable and a meta-stable phase
in the infinite volume limit ($L\rightarrow\infty$),
the random walker staying near the favored corner F=($1,L-1$) and the 
unfavored corner U=($L-1,1$) respectively.
Above $h_c$ there exists only one phase, the random walker 
always staying around F.
This is also reflected in the spectrum of the hamiltonian (\ref{maeqmatrix}),
since one has two ground states for $L\rightarrow\infty$ and $h<h_c$ 
and only one for $h>h_c$.

It is also possible to measure the flip time
from one phase to the other using Monte Carlo simulations.
At the spinodal point, one observes a discontinuity in the passage time
from the corner U to the corner F.
At and above $h_c$, the average time taken to reach the corner F grows linearly with $L$.
However, there is a discontinuity in the proportionality constant as
one approaches $h_c$ from above. The linearity above $h_c$ can be explained
by a mean-field argument, the linearity at $h_c$ cannot.
The time taken to reach the corner U and the time taken to reach the corner F for $h<h_c$
grow exponentially with $L$.

Using Monte Carlo simulations, one can also investigate the 
movement of the random walker starting from specific initial conditions.
Analysing the mean path, one again identifies two time scales away from the spinodal point,
and at the spinodal point, one again finds universal behavior.

All the results found here for the RWT model are in accordance with the
results found for the three-state and the asymmetric diffusion models.
But the detailed analysis of the spectrum of the time evolution
operator now suggests a link between the time-scales found in the spectrum,
the time-scales found in the flip times, and the scales
one observes in the movement of the random walker. 
This was not observed before. 

The random walker on a triangle can be simplified to a 
random walker on a right angle (RWRA) model which can be handled analytically
while still possessing a meta-stable phase. 
This is done taking $b=0$ 
and keeping only the points $(j,k)\in {\cal T}_L$
with $j=1$ or $k=1$.
That is, we consider only the movement along the
$j$- and $k$-axes, where macroscopic jumps occur.
In an infinitesimal time interval $\d t$ the following processes
can occur with a probability $p$:
\begin{eqnarray}
\label{sutoy}
(j,1)\rightarrow (j-1,1)\qquad &\mbox{if $j>1$ with }p=a(1+h)\d t
\nonumber\\
(j,1)\rightarrow (L-1,1)\qquad &\mbox{with }p=c(1-h)\d t
\nonumber\\
(1,k)\rightarrow (1,k-1)\qquad &\mbox{if $k>1$ with }p=a(1-h)\d t
\nonumber\\
(1,k)\rightarrow (1,L-1)\qquad &\mbox{with }p=c(1+h)\d t
\end{eqnarray}
The RWRA model is essentially a one-dimensional (albeit non-local)
random walker model.
Again we limit our investigation to the case $c=1/2$
(this fixes the unit of time) 
and $a\equiv a+b=1/2$. 
We checked that taking $a$ different from $1/2$ does not change 
the physics of the RWRA model.
One is able to calculate exactly
a characteristic equation for the eigenvalues of the
time evolution operator, the free energy functional
and the flip times.
The spinodal point is at $h_c=1$.
Since all rates in (\ref{sutoy}) have to be positive one can study
with the RWRA model only
the regime below the spinodal point. 
The results are given in appendix C.

The paper is organized as follows.
In the next section, we present our results on the spectrum of the hamiltonian
of the RWT model.
We identify two time scales, one of which vanishes at the spinodal point.
At this point, the low-lying excitations acquire a universal $L^{-1}$ dependence
yielding $z=1$. We also show that
finite-size scaling relations hold.

In section 3, we compute the FEF of the RWT model and present
our Monte Carlo results on the passage times from the favored corner F to the
unfavored corner U and vice versa. 
Two scales can be measured and related
to the depths of the minima of the FEF and to gaps 
in the spectrum of $H$ found in section 2.

In section 4, we give results on the movement of the random walker that
mirror the scales found for the spectrum (section 2).
At the spinodal point, one finds a non-trivial universal 
behavior that also reflects the dynamical exponent $z=1$. 
Section 5 contains our conclusions.

In appendix A, we give some simple results for the movement of the random walker
for $h$ above the spinodal point.
In appendix B we explain the connection between the flip times and 
the spectra of models with absorbing configurations.
In appendix C we investigate the random walker on a right angle model.
%%%%%%%%%%%%%%%%%%%%%%%%%%%%%%%%%%%%%%%%%%%%%%%%%%%%%%%%%%
%
%
\def\figuremovetwo{
\begin{figure}[tb]
\setlength{\unitlength}{1mm}
\def\setl{\setlength\epsfxsize{10cm}}
\begin{picture}(80,82)
% plot "dist_0.250_0.250_0.50000_100.mod"t"","dist_0.250_0.250_0.50000_400.mod"t""
\put(20,0){
        \makebox{
                \setl
                \epsfbox{move2.epsf}}
        }
\put(20,65){\makebox{$s(t)$}}
\put(127,1){\makebox{$t/L$}}
\end{picture}
\caption{Average distance $s(t)$ from starting point U=$(L-1,1)$.\protect\\ 
         Data from
         simulation of the RWT model for
         $a=\frac{1}{4}$, $h=\frac12>h_c$.
	The solid curve represents the data for $L=100$, the 
	dashed for $L=400$.
	The scaling variable is $t/L$.
\label{figmovetwo}}
\end{figure}
}
\def\figurespectrum{
\begin{figure}[tb]
\setlength{\unitlength}{1mm}
\def\setl{\setlength\epsfxsize{10cm}}
\begin{picture}(80,82)
\put(20,0){
        \makebox{
                \setl
                \epsfbox{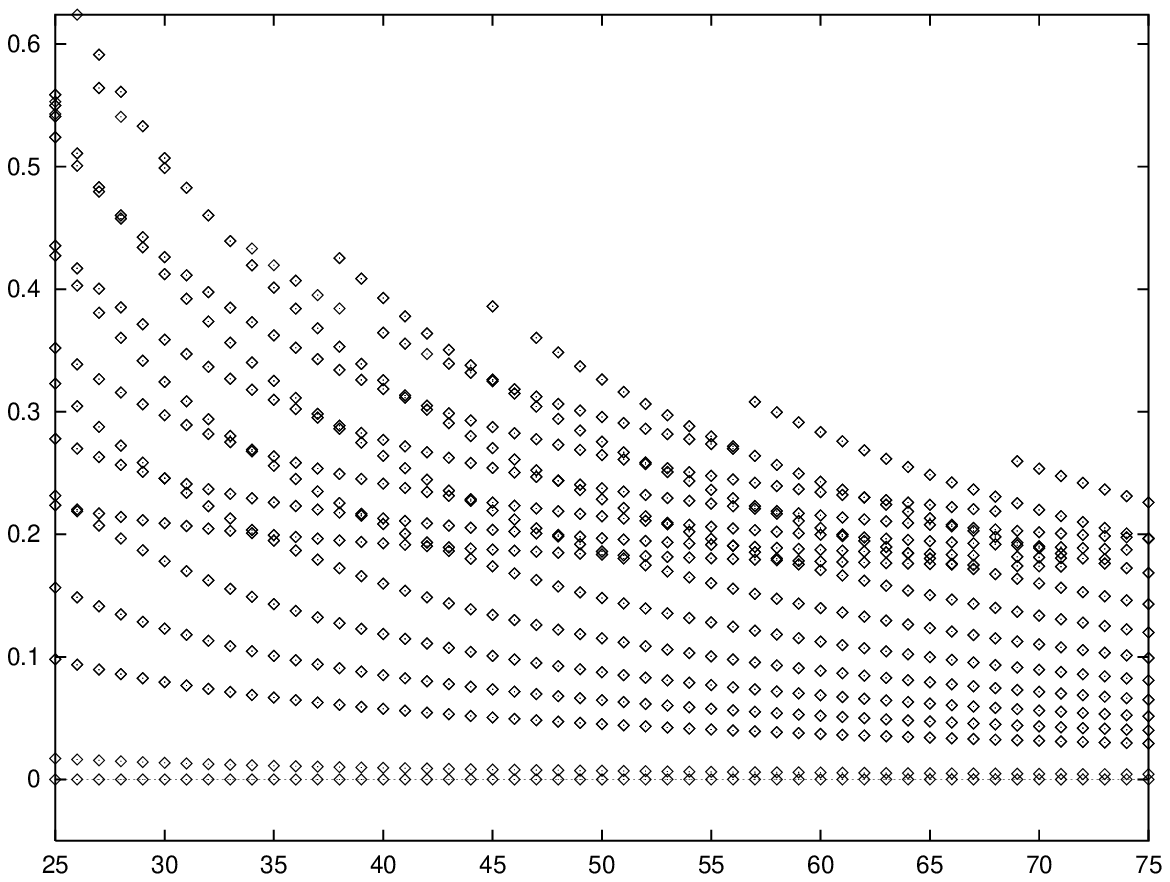}}
        }
\put(80,48){\makebox{$E_1=\exp(-\omega\,L)$}}
\put(80,58){\makebox{$E_0=0$}}
\put(120,11){\vector(1,0){4}}
\put(120,26){\vector(1,0){4}}
\put(125,10){\makebox{$m_1$}}
\put(125,25){\makebox{$m_2$}}
\put(14,65){\makebox{${\rm Re}(E_i)$}}
\put(126,-1){\makebox{$L$}}
%\put(25,0){\makebox{Numerical diagonalization of ${\cal H}$: $a=0.25$, $h=0.3$}}
\end{picture}
\caption{Dependence of the spectrum on system size. The data is
         from numerical diagonalization for $a=\frac{1}{4}$, $h=0.3$.
	 Shown are the real parts of the eigenvalues of $H$.
	 The gaps $m_1$ and $m_2$ extrapolate to finite values as seen in
	 figure \protect\ref{figmasses}.
\label{figspec}}
\end{figure}
}
\def\figuremasses{
\begin{figure}[t]
\setlength{\unitlength}{1mm}
\def\setl{\setlength\epsfxsize{10cm}}
\begin{picture}(80,82)
\put(20,0){
        \makebox{
                \setl
                \epsfbox{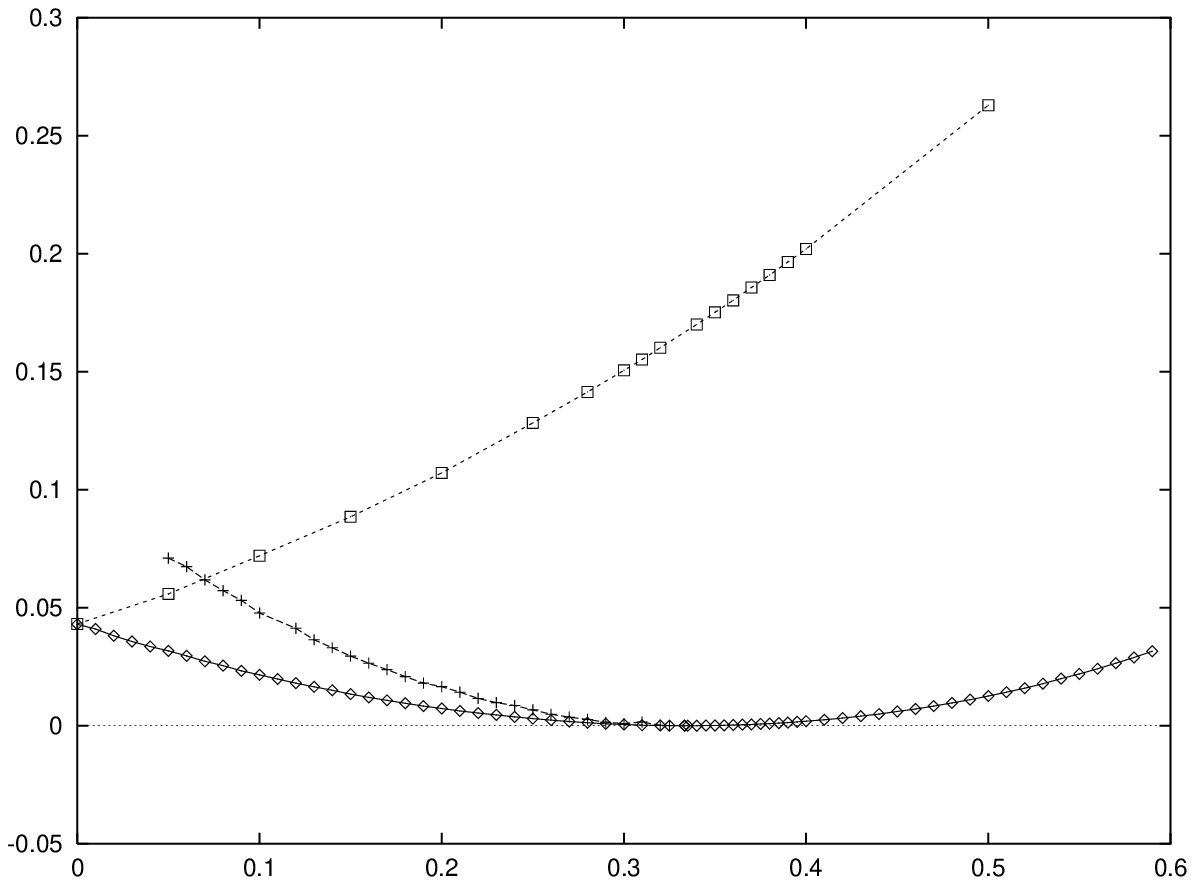}}
        }
\put(106,18){\makebox{$m_1$}}
\put(38,15){\makebox{$m_1$}}
\put(51,20){\makebox{$\omega$}}
\put(80,49){\makebox{$m_2$}}
\put(80,8){\line(0,1){3.5}}
\put(79,4){\makebox{$h_c$}}
\put(21,65){\makebox{$m$}}
\put(127,0){\makebox{$h$}}
%\put(25,0){\makebox{From numerical diagonalization of ${H}$: $a=0.25$}}
\end{picture}
\caption{Length scale $\omega$, and upper and lower gaps $m_1$ and $m_2$ 
         for $a=\frac{1}{4}$. 
	 The values given are extrapolated from data of
         numerical diagonalization of $H$ up to $L=220$.
\label{figmasses}}
\end{figure}
}
\def\figuremove{
\begin{figure}[tb]
\setlength{\unitlength}{1mm}
\def\setl{\setlength\epsfxsize{10cm}}
\begin{picture}(80,82)
% plot "dist_0.250_0.250_0.30000_100.mod"t"","dist_0.250_0.250_0.30000_600.mod"t""
\put(20,0){
        \makebox{
                \setl
                \epsfbox{move.epsf}}
        }
\put(20,65){\makebox{$s(t)$}}
\put(127,1){\makebox{$t/T_{\rm short}$}}
\end{picture}
\caption{Average distance $s(t)$ from starting point U=$(L-1,1)$.\protect\\
         Data from
         simulation of the RWT model for
         $a=\frac{1}{4}$, $h=0.3<h_c$.
	The solid curve represents the data for $L=100$, the 
	dashed for $L=600$.
\label{figmove}}
\end{figure}
}
\def\figuref{
\begin{figure}[tb]
\setlength{\unitlength}{1mm}
\def\setl{\setlength\epsfxsize{10cm}}
\begin{picture}(80,82)
\put(20,0){
        \makebox{
                \setl
                \epsfbox{fef.epsf}}
        }
\put(20,65){\makebox{$f_L$}}
\put(127,1){\makebox{$\M$}}
\end{picture}
\caption{The free energy functional $f_L(\M)$ for $a=\frac14$, $L=100$ and
$h=0.0, 0.1, 0.2, 0.3333, 0.4$.
The values are shifted by the value $f_L(-1)$.
The data is generated by %from the results 
numerical diagonalization of the hamiltonian.
Since the precision is limited to $10^{-12}$ the curves for
larger $h$ cannot be fully determined.
\label{figf}}
\end{figure}
}
\def\figureflip{
\begin{figure}[tb]
\setlength{\unitlength}{1mm}
\def\setl{\setlength\epsfxsize{10cm}}
\begin{picture}(80,82)
\put(20,0){
        \makebox{
                \setl
                \epsfbox{flip.epsf}}
        }
\put(20,65){\makebox{$T$}}
\put(127,1){\makebox{$L$}}
\end{picture}
\caption{The flip times $T_{\rm short}(L)$ and $T_{\rm long}(L)$
for $a=\frac14$ and $h=0.1$.
The fitted lines are given by 
$50\exp(0.2\,L)$ and $500\exp(0.052\,L)$.
\label{figflip}}
\end{figure}
}
\def\figuremovethree{
\begin{figure}[tb]
\setlength{\unitlength}{1mm}
\def\setl{\setlength\epsfxsize{10cm}}
\begin{picture}(80,82)
% plot "dist_0.250_0.250_0.33333_100.mod"t"","dist_0.250_0.250_0.33333_400.mod"t""
\put(20,0){
        \makebox{
                \setl
                \epsfbox{move3.epsf}}
        }
\put(20,65){\makebox{$s(t)$}}
\put(127,1){\makebox{$t/L$}}
\end{picture}
\caption{Average distance $s(t)$ from starting point U=$(L-1,1)$.\protect\\
         Data from
         simulation of the RWT model for
         $a=\frac{1}{4}$, $h=h_c$.
	The solid curve represents the data for $L=100$, the 
	dashed for $L=400$.
\label{figmovethree}}
\end{figure}
}
\def\figuremovefour{
\begin{figure}[tb]
\setlength{\unitlength}{1mm}
\def\setl{\setlength\epsfxsize{10cm}}
\begin{picture}(80,82)
% plot "dist_0.250_0.250_0.30000_100.mod"t"","dist_0.250_0.250_0.30000_200.mod"t"","dist_0.250_0.250_0.30000_400.mod"t"","dist_0.250_0.250_0.30000_600.mod"t""
\put(20,0){
        \makebox{
                \setl
                \epsfbox{figmove4.epsf}}
        }
\put(15,65){\makebox{$s(t)L$}}
\put(123,1){\makebox{$t$}}
\end{picture}
\caption{Average distance $s(t)$ from starting point U=$(L-1,1)$.\protect\\
         Data from
         simulation of the RWT model for
         $a=\frac{1}{4}$, $h=0.3$ and system sizes $L=100, 200, 400, 600$.
	The scaling variable is $t$.
\label{figmovefour}}
\end{figure}
}
%
%
%%%%%%%%%%%%%%%%%%%%%%%%%%%%%%%%%%%%%%%%%%%%%%%%%%%%%%%
%
\def\tablecriticallevel{
\begin{table}[tb]
\caption{Excited levels of $H$ at $h_c$:
         $E_j=a\frac{k_{j}}{L}$. Complex excitations
	 come in conjugated pairs.
\label{tabcritlev}}
\begin{eqnarray}
\begin{array}{@{}llll} \br
\mbox{level} & \mbox{Re($k^{(1)}_j$)} & \mbox{Im($k^{(1)}_j$)} & \mbox{Im($k^{(1)}_j$)}/12.81\\ \mr
1 & 1.7080 & 0       & 0               \\
%2 & 8.4980 & 12.7663 & 1              \\                   %0.997
2 & 8.4980 & 12.7663 & 0.997           \\
%4 & 10.368 & 25.771  & 2.019          \\                   %2.012
4 & 10.368 & 25.771  & 2.012           \\
%6 & 11.504 & 38.548  & 3.020          \\                   %3.009
6 & 11.504 & 38.548  & 3.009           \\
%8 & 12.324 & 51.241  & 4.014          \\                   %4.000
8 & 12.324 & 51.241  & 4.000           \\
%10 & 12.97 & 63.89   & 5.005          \\                   %4.988
10 & 12.97 & 63.89   & 4.988           \\
%12 & 13.5 & 76.52    & 5.994          \\ \br               %5.974 
12 & 13.5 & 76.52    & 5.994          \\ \br               %5.974 
\end{array}
\nonumber
\end{eqnarray}
\end{table}
}
%%%%%%%%%%%%%%%%%%%%%%%%%%%%%%%%%%%%%%%%%%%%%%%%%%%%%%%%%%%%%%%%%%%%%%%%%%%%%%%%%%%%
\section{Spectrum of the hamiltonian\label{sectionspec}}
\figurespectrum
%
%The RWT model corresponding to a three-state model of $L$ sites has only
%$L(L-1)/2$ configurations. Therefore it becomes possible to study the spectrum
%of its time evolution operator $\H$ and to extrapolate 
%the finite size data 
%to the large $L$ limit.
%For the three-state model this could only be done for the first excitation
%and with a less precise extrapolation from up to 11 sites \cite{us}.
%
Using a modified version of the Arnoldi algorithm \cite{uli}, 
we have calculated the
low-lying eigenvalues (up to the 30th level) of the time evolution
operator $\H$ for $L\leq 220$,
various values of $h$,
and $a=0.001,\, 0.1,\, 0.25,\, 0.4$.
Apart from the ground state ``energy'' $E_0(L)\equiv 0$,
the further eigenvalues of the non-hermitian matrix
$\H$ are positive, or complex with positive real parts.
Since the hamiltonian matrix has real entries only,
complex eigenvalues come in conjugated pairs.
Typical results for the real part of the spectrum
(for $a=0.25 , h=0.3$) are shown in figure \ref{figspec}.
One identifies two bands of excitations leading to the gaps 
(time scales) $m_1$ and $m_2$ in the large $L$ limit.
Note that the excitations correspond to energy gaps, since the
ground state energy of $\H$ is zero for any system size.

Values for the large $L$ limit were extrapolated from
data for finite
$L$ using the Bulirsch-Stoer algorithm \cite{BST}.

The most important observation is that there exists a
critical spinodal point at $h_c=\frac{a}{1-a}$ (see figure 3).
At this point,
the system loses one of its time scales ($m_1=0$)
and acquires a band of excitations with a universal $L^{-1}$
dependence.

The results for the spectrum of
$\H$ can be summarized as follows:
\figuremasses
\begin{itemize}
\vspace{6pt}
\item
{\it first excitation:}\\
For $h<h_c$ the first excited energy level approaches zero as 
\begin{equation}
\label{lev1}
E_1\propto \exp(-\omega L)
\end{equation}

This excitation is real, the higher excitations are complex conjugated pairs.
Thus, in the limit $L\rightarrow\infty$, there are two ground states. Linear combinations
of the eigenvectors give two phases \cite{BeGr}.
For $h=0$ they are equally stable, while for $h>0$
one of them is stable, and the other is meta-stable.
The two phases correspond to minima of the free energy functional
at $\M=1$ and $\M=-1$ (see section 3).
The meta-stable phase vanishes at the spinodal point $h_c$.
Above $h_c$, there is only one ground state for $L\rightarrow\infty$,
i.e. none of the excited levels approaches zero.
\vspace{6pt}
\item
{\it higher excitations:}\\
Above $E_0$ and $E_1$, 
there are two bands of complex eigenvalues of the form
\begin{eqnarray}
E^{(1)}_{j}(L)&=m_1+k^{(1)}_j/L+\cdots
\label{lowspec}\\
E^{(2)}_{j}(L)&=m_2+k^{(2)}_j/L+\cdots
\;\;.
\label{highspec}
\end{eqnarray}
The gaps $m_1$ and $m_2$ are real, whereas the constants
$k^{(i)}_j$ have imaginary parts, 
such that one has complex conjugated pairs of eigenvalues. 
One finds that $m_1$ is proportional to $\omega$.
%
%\begin{equation}
%m_1\propto\omega
%\end{equation}
%
%with a proportionality constant 0.57(4) 
For $a=0.25$.
we find $m_1=0.57(4)\;\omega$.

For $h=0$ one has $m_1=m_2$ and the constants $k^{(1)}_j=k^{(2)}_j=k$
are independent of~$j$, i.e. 
the $1/L$ term is the same for all excitations
in contrast to Eqs.(\ref{lowspec}) and (\ref{highspec}) for $h>0$.
There are two ground states
(compare Eq.(\ref{lev1})), corresponding to equally stable phases.

For $h>h_c$
the excitations again come in the form of two bands as in Eqs.(\ref{lowspec})
and (\ref{highspec}).
\vspace{6pt}
\item
{\it lower gap:}\\
For $h\rightarrow h_c$, the
lower gap $m_1$ vanishes, as does $\omega$, which is proportional
to $m_1$.
For $h<h_c$, the gap is given by $m_1=\lim_{L\rightarrow\infty}E_2(L)$.
Above $h_c$, the first excitation does not give a second
ground state for $L\rightarrow\infty$ and one finds a non-zero gap 
$m_1=\lim_{L\rightarrow\infty}E_1(L)$. The gap $m_1$ given by
\begin{eqnarray}
m_1=&0.42(2)\; (h_c-h)^x \;\;\;&\mbox{ for }h<h_c\nonumber\\
\label{gapsexp}
m_1=&0.46(2)\; (h-h_c)^x &\mbox{ for }h>h_c
\end{eqnarray}
where the numbers are given for $a=0.25$.
Note that these equations are valid for any $h$ and not only in the vicinity of $h_c$.
One finds an exponent $x=2.00(1)$ for
$h<h_c$ and $h>h_c$.
Extrapolated values for the $m_i$ are shown in figure \ref{figmasses}.

The position of the minimum of the parabola Eq.(\ref{gapsexp}) determines 
the spinodal point at $h_c=\frac{a}{1-a}$ with a precision of two digits.
\vspace{6pt}
\item
{\it upper gap:}\\
The scale $m_2$ increases monotonically with $h$ over the entire range $0\leq h <1$. One finds
\begin{equation}
m_2=0.48(2) h^{1.22(2)}+0.043(2) 
\end{equation}
for $a=0.25$.

\tablecriticallevel

\vspace{6pt}
\item
{\it lower band of excitations at $h=h_c$:}\\
At the spinodal point, the lower band is of the following form:
\begin{eqnarray}
\label{critspectrum}
E_0=0& \nonumber\\
E_1=a\frac{k^{(1)}_1}{L}+\frac{\mbox{const.}}{L^{3/2}}+\cdots  &\mbox{real}
\label{eqz}\\
E_{j}=a\frac{k^{(1)}_{j}}{L}+\frac{\mbox{const.}}{L^{3/2}}+\cdots
\quad &j\geq 2,\mbox{ complex conjugated pairs,} \nonumber
\end{eqnarray}
with the extrapolated values $k_j^{(1)}$ given in table \ref{tabcritlev}.
One has universality:
apart from the normalization constant $a$
(``speed of light''), the leading terms are the same
for any critical point $h_c(a)=\frac{a}{1-a}$.
Extrapolation of the $k^{(1)}_i$ for $a=$0.25, 0.4, 0.1
gives
the same result with 5 digits precision.
At the spinodal phase transition,
the dynamical critical exponent \cite{dynexp}
is $z=1$, as was found for the three-state diffusion model
where, however, this was only seen for the first excitation
\cite{us}.\\
The imaginary parts of the
numbers $k^{(1)}_j$ follow (see table \ref{tabcritlev})
\begin{equation}
%E_{2j}=\overline{E_{2j+1}}=E_1(L)+\mbox{const.}(j/(j+1)+{\rm i}\;j)L^{-1}
{\rm Im}(k^{(1)}_j)=12.81(1) \;j\; .
\end{equation}
This equidistant spacing comes as a surprise and we do not know its reason.
For quantum spin chains related to a conformal field theory
one expects regularities (conformal towers) for the real parts of the $k^{(1)}_j$.
However, we could not detect any such structure looking at the real part
of the spectrum. This should be a result of the reduced state space of the RWT
model.
Because of the large configuration space of the
full three-state model, a similar analysis was not possible there.

For the higher band of excitations at the spinodal point,
one again has the form of
Eq.(\ref{highspec}).
\vspace{6pt}
\item
{\it finite-size scaling analysis:}\\
Finite-size scaling relations, which describe
the behavior of finite quantum spin chains in the vicinity
of an (equilibrium) critical point of the infinite system \cite{FiSiSc},
apply also to the (non-equilibrium) RWT model at $h_c$.
Consider the curves
\begin{equation}
F_L:\;h\rightarrow L\, {E_1(h,L)} \;\;.
\end{equation}
The critical point $h_c$ can be determined
from extrapolation of the crossing points 
$h_{\mbox{\footnotesize cross}}(L)$ of the curves $F_L$ and $F_{L-1}$.
Indeed one finds
\begin{equation}
\lim_{L\rightarrow\infty} h_{\mbox{\footnotesize cross}}(L)=0.3333(2)=h_c
\end{equation}
for $a=0.25$. This is the most precise numerical determination of $h_c$.

Following the finite-size scaling hypothesis, one can also determine the
critical exponent $x$ in a way different from that of Eq.(\ref{gapsexp}) \cite{FiSiSc}.
From the available data up to $L=220$, one again finds $x=2.0(2)$. 
%For an equilibrium system, $x$ would be the critical exponent of the
%correlation length.
\vspace{6pt}
\end{itemize}
In the following sections, we discuss further properties of the RWT model
that can be related to the scales of the spectrum and to the
universality at the critical point.
The scale $\omega$ can be linked with
flip times and the stationary \FEF\ (see next section).
In section 4,
we present
dynamic quantities related to the
properties of $\H$, specifically to the gaps $m_i\/$.
%
%%%%%%%%%%%%%%%%%%%%%%%%%%%%%%%%%%%%%%%%%%%%%%%%%%%%%%%%%%%%%%%%%%%%%%%%%%%%%%%%%%%%
\section{Free energy functional and flip times between phases
\label{sectionev} \label{sectionflip}}
\figuref
\figureflip

We have investigated the behavior of the stationary probabilities
$P_0(j,k)$ for different $h$.
These probabilities were determined from the stationary state as
computed by numerical diagonalization of the 
hamiltonian (see section 2).
First, we remark that
for $h=0$ all eigenstates are also
eigenstates of the $CP$ operation.
We find that
the eigenvectors corresponding to the complex conjugated pairs of
eigenvalues $E^{(1)}_{j}$ of the lower band of excitations 
(\ref{lowspec})
are $CP^+$ for $j$ even
and $CP^-$ for $j$ odd.

One also finds that,
for any $h$ and finite $L$, the eigenvector $\Ps$ with eigenvalue $E=0$ 
is the only vector that directly corresponds
to a probability distribution, i.e.
all its entries are positive.
In the case $h=0$, $\Ps$ has $CP^+$ symmetry.

Hence, for $h=0$, the free energy functional 
$f(\M)$, defined in Eq.(\ref{deffreeenergy}),
is symmetric under the inflection $\M\rightarrow -\M$.
(From the data, one finds that the free energy functional
is a well defined limit
as was already known for the asymmetric exclusion model 
and the three-state diffusion model \cite{us}.)
There are minima at $\M=-1$ and $\M=1$ and
a maximum at $\M=0$.
The minima correspond to the two equally stable
phases (see figure~\ref{figf}).
This is basically the same situation as for the $CP$-symmetric
three-state model in the ``low-temperature'' phase.

For $0<h<h_c$, one has a global
minimum at $\M=-1$ and a further, shallower minimum at $\M=1$
(see figure \ref{figf}).
This reflects the fact that one of the phases
becomes meta-stable.
One can define two scales:
\begin{eqnarray}
\label{dendiffmass}
\nu_{\rm large}=f_{\rm max}-f(-1)
\nonumber\\
\nu_{\rm small}=f_{\rm max}-f(1)\; ,
\end{eqnarray}
where $f_{\rm max}$ is the maximum value of $f(\M)$.
We shall relate these scales to the time
the system needs to flip from one minimum to the other.
At $h_c$, the second minimum of $f$
(and hence the meta-stable phase) vanishes.
This cannot be seen in figure \ref{figf} but can be inferred
from the results for smaller lattice sizes.
For $h\geq h_c$, the free energy has only one minimum at $\M=-1$.
%and one can only define $\nu_{\rm large}$.

\vspace{6pt}

Analysis of the eigenvectors of $\H$
that become
the steady states in the large $L$ limit,
and the relation of the RWT to the three-state diffusion
model, have given the result that below $h_c$
there are two distinct phases
with order parameter values
$\M=1$ and $\M=-1$.
Above $h_c$, however, there is only one phase: the walker stays
near the favored corner F of ${\cal T}_L$.
This can be verified with Monte Carlo simulations.
In a way similar to that for three-state diffusion model \cite{orig,us},
one can measure the time for the RWT model to flip from
one phase to the other.
One finds that, for large $L$ and $h<h_c$, in the neighbourhood of a corner 
F=$(1,L-1)$ or U=$(L-1,1)$,
the random walker moves through the configurations on a
shorter time scale than is needed to get from one corner to the other.
This allows one to measure the flip times $T_{\rm long}$ from favored to unfavored
and $T_{\rm short}$ from unfavored to favored phase
as average first passage times
from F to U and U to F respectively.

With $c=\frac12$ and $a+b=\frac12$, the unit of time is
defined such that, on average, the random walker moves
one step (\ref{toy1}) per unit.
From the 
simulations, one finds that, in leading order, the flip times are given by
\begin{eqnarray}
\label{flipmasslong}
T_{\rm long}&\propto\exp(\mu_{\rm long}\,L)\qquad\;\,\mbox{for any }h\\
\label{flipmassshort}
T_{\rm short}&\propto\left\{\begin{array}{ll}
                              \exp(\mu_{\rm short}\,L)&\mbox{ for }h<h_c\\
			      L&\mbox{ for }h\geq h_c\;.
				\end{array}
			\right.
\end{eqnarray}
For $h<h_c$, the time intervals the random walker spends alternately near 
either corner
increase exponentially
with $L$. The fraction of time in the meta-stable region
decreases exponentially.
In this sense, for $h<h_c$, there are a meta-stable and a stable phase.
Results of the measurement for $a=0.25$ and $h=0.1$ are presented in figure \ref{figflip}.

For $h>h_c$, the proportionality constant
in Eq.(\ref{flipmassshort})
can be calculated by a simple argument, see appendix A:
\begin{eqnarray}
\label{fliphigh}
T_{\rm short}^{h>h_c}&=\frac{L}{h+a(1-h)}=\frac{1+h_c}{h+h_c}L\; .
\end{eqnarray}

In contrast, the appearance of a linearly growing flip time at $h=h_c$
is non-trivial.
One finds the form
\begin{eqnarray}
\label{flipcrit}
T_{\rm short}^{h_c}=\frac{L}{a} - \frac{\gamma L^{1/2}}{a} + \mbox{const.}\;,
\end{eqnarray}
where $\gamma=0.57(6)$ is a positive constant and independent of $a$
to a precision of about $10\%$.
Up to the normalization $a$,
the short flip time is, in leading order, a universal function of $L$ 
as are the leading terms of the lower spectrum Eq.(\ref{critspectrum}).
The appearance of a universal linear flip time is linked
to the universal $1/L$ spectrum of the hamiltonian, compare Eq.(\ref{critspectrum}).
One can ask how general this universal behavior is.
Similar investigations for other models remain to be done.

Note that from (\ref{fliphigh}) and (\ref{flipcrit}) one finds that in leading order
\begin{equation}
\lim_{h\rightarrow h_c} T_{\rm short}^{h>h_c} = 2 T_{\rm short}^{h_c}\;.
\end{equation}
Hence there is a discontinuity in the flip time $T_{\rm short}^{h_c}$.
At the spinodal point,
$T_{\rm short}^{h_c}$ has a special $L$ dependence,
essentially different from its behavior above or below this point.

The values of 
$\nu_{\rm small}$ and $\nu_{\rm large}$
in (\ref{dendiffmass}), determined from
numerical diagonalization of $\H$ for $L\leq 140$,
and measurement of
the flip times,
up to $L=200$ for $T_{\rm short}$ and $L=40$ for
$T_{\rm long}$, yield the following result:
\begin{eqnarray}
\mu_{\rm short}=\nu_{\rm small}\label{flipeq}
\quad&\mbox{  for }h<h_c\nonumber\\
\mu_{\rm long}=\nu_{\rm large}&\mbox{  for any }h\; .
\label{flipmasses}
\end{eqnarray}
As an example for $h=0.1$ and $a=0.25$, we find the values $\mu_{\rm short}=0.052(3)$ 
and $\mu_{\rm long}=0.20(2)$ (see Fig.\ref{figflip})
which is to be compared
with $\nu_{\rm small}=0.053(2)$ and
$\nu_{\rm large}=0.17(2)$ (see Fig.\ref{figf}).
For $h=0$ and $a=0.25$ the equations (\ref{flipmasses})
can be checked to three digits;
the precision decreases
with increasing $h$ to about one digit at $h=0.3$.
(Finite-size corrections to
(\ref{dendiffmass}), (\ref{flipmasslong}) and (\ref{flipmassshort}) 
become more important for
higher $h$ due to
crossover effects to the linear regime for 
$h\geq h_c$.)

Equation (\ref{flipmasses}) is interesting because
it relates (steady state) properties of the
free energy functional $f(\M)$
with (dynamical) flips between phases.
This is unexpected for a non-local model
without detailed balance.
For the RWRA model (appendix C) an analogous link can be made analytically.
Also, for the three-state diffusion model,
similar observations, albeit with less precision, were
made in \cite{us}.

One also finds that $\mu_{\rm short}=\omega$
(recall that $E_1\propto\exp(-\omega L)$ is
the first excitation of the hamiltonian, see Eq.(\ref{lev1})).
For
$h=0.1$ and $a=0.25$ the value of
$\omega$ from numerical diagonalization is $0.048(3)$.
The reason for this connection,
together with an analogous expression for $\mu_{\rm long}$,
is given in appendix B, where we
consider modified models with absorbing configurations.

%%%%%%%%%%%%%%%%%%%%%%%%%%%%%%%%%%%%%%%%%%%%%%%%%%%%%%%%%%%%%%%%%%%%%%%%%%%%%%%%%%%%
\section{Time scales
and criticality in the movement of
the random walker\label{sectionmove}}
To find macroscopic physical quantities reflecting
the time scales of the spectrum and the critical behavior at $h_c$,
we investigated the 
movement of the random walker starting from the specific
initial positions U$=(L-1,1)$ (unfavored corner) or 
F$=(L-1,1)$ (favored corner).
In the following we concentrate on the 
time dependence of the order parameter coordinate.
Defining the average order parameter distance 
from the starting point as
\begin{equation}
\label{defdist}
s(t)=\frac{\left|<\M(t)-\M(0)>\right|}{2} \;,
\end{equation}
the start configuration corresponds to $s=0$ and
the opposite corner of the triangle ${\cal T}_L$ to $s=1$.

\subsection*{Start at unfavored corner}
\figuremovefour
\figuremove
First, we consider the random walker starting at the 
point U.
We shall see that the movement from the unfavored to the favored corner
is linked with the lower excitations of the hamiltonian.

Typical pictures of the walker's position $s(t)$ are given in figures
\ref{figmovefour}, \ref{figmove}, \ref{figmovethree} and \ref{figmovetwo}.
In figure \ref{figmovefour} we show the product
$L\cdot s$  as a scaling function of $t$, in figure
\ref{figmove} the distance $s$ as a scaling function of $t/T_{\rm short}$ (both for 
$h$ below the critical point),
and in figures
\ref{figmovethree} and \ref{figmovetwo} the distance
$s$ as a scaling function of $t/L$ 
(at and above
the critical point respectively).
One observes the following properties of the scaling functions:
\begin{itemize}
\vspace{6pt}
\item
$h<h_c$:\\
The random walker
stays near U for an average time exponentially increasing
with $L$ (Eq.(\ref{flipmassshort})).
For large $L$,
the path near the unfavored corner (start)
is described by the 
scaling function
\begin{equation}
\label{beginlow}
L s(t)\propto (1-\exp(-\lambda t))\; .
\end{equation}
Here we find $\lambda=1.8(2)\cdot m_1(h)$ for $a=0.25$.
Due to the distribution of flip times for finite $L$ the walker leaves this
regime at different times (figure \ref{figmovefour}).

For the approach to the favored corner, one observes a different scaling
(figure \ref{figmove}). 
From the data, one finds that
$s\/$ is given by
\begin{equation}
\label{endlow}
s(t,L)=s_\infty-\exp(-\lambda \frac{t}{T_{\rm short}(L)})\; ,
\end{equation}
where 
\begin{equation}
s_\infty=\lim_{t\rightarrow\infty}s(t)=1-\frac{\mbox{const.}}{L}+\cdots \;\;.
\end{equation}
In Eq.(\ref{endlow}), the constant $\lambda=1.15(3)$ is independent of $a$.
The approach to the favored corner scales with 
$t/T_{\rm short}$.
In the scaling of figure \ref{figmove}, the initial path (Eq.(\ref{beginlow})
and figure \ref{figmovefour})
of the
random walker cannot be seen because it is reduced to a point.
\figuremovethree
\vspace{6pt}
\item
$h=h_c$:\\
We consider the scaling regime $t\rightarrow\infty$ and $L\rightarrow\infty$,
taking $t/L$ fixed.
For $t<T_{\rm short}/2=L/2a$, the distance from the starting point 
increases in leading order linearly 
with time. We find
\begin{equation}
\label{begincrit}
s(t,L)=\frac1L\left( \frac{a}{2}t+\delta\sqrt{t}+ \mbox{const.}+\cdots\,\right)\; ,
\end{equation}
where $\delta=0.17(2)$ independent of $a$.

For large $t> T_{\rm short}/2$, the random walker approaches $s=1$
exponentially, as in the case $h\neq h_c$. The scale, however, is
the size of the system
\begin{equation}
\label{endcrit}
s(t,L)=s_\infty-(s_\infty-a)\exp(-\kappa  (a\frac{t}{L}-\frac{1}{2})) \; ,
\end{equation}
with $\kappa=1.7(1)$ independent of $a$.
Typical data is shown in figure \ref{figmovethree}.

At the critical point, the initial path of the random walker
Eq.(\ref{begincrit}) 
is universal and linear in leading order.
This reflects again the universal $1/L$ spectrum found for the critical point
(\ref{critspectrum}).
It is remarkable that 
the exponential scale of Eq.(\ref{endcrit}) is also universal
(but not the scaling function).
Up to the time normalization (as in section 2),
they do not depend on $a$
and scale with $tL^{-1}$. 
\vspace{6pt}
\figuremovetwo
\item
$h>h_c$:\\
Again we study the scaling $t/L$ fixed taking $t$ and $L$ to infinity.
Starting from the unfavored corner U, at first
(for $t<T_{\rm short}$) and in leading order the random walker
follows a mean path determined by its velocity vector
(see appendix A). The corrections are exponential in $t$. We find
\begin{equation}
\label{beginhigh}
s(t,L)=ah\frac{t}{L}-\frac{\zeta (1-\exp(\xi t))}{2L}\; .
\end{equation}
With increasing $h$,
the constant $\zeta$ decreases and $\xi$ increases and thus the corrections become smaller,
consistent with the explanation of appendix A. 

For $t>T_{\rm short}$,
one again finds an exponential approach to the final value $s=1$
(see figure \ref{figmovetwo}). One has
\begin{equation}
\label{endhigh}
s(t,L)=s_\infty-\eta \exp(-(m_1+\frac{\chi}{L})(t-T_{\rm short}))\,,
\end{equation}
where $T_{\rm short}$ is given by (\ref{fliphigh}).
The scale is the gap $m_1$ from the spectrum of the time evolution operator.
Following 
appendix A, one finds the prefactor
\begin{equation}
\label{eta}
\eta=\frac1{1+h/h_c}\; .
\end{equation}
\end{itemize}
\subsection*{Start at the favored corner}
We have also studied the path of the random walker starting from F.
In that case the critical point $h_c$ does not play a special role.
From the simulations, one finds
\begin{equation}
\label{beginfavabs}
s(t,L)\propto \frac{1-\exp(\lambda t)}{L}\; ,
\end{equation}
where $\lambda=1.25(5)\cdot m_2(h)$ for $a=0.25$.
This equation is similar to Eq.(\ref{beginlow}), but
note that
in this case the movement is related to the time scale $m_2$ that describes
a faster movement than the time scale $m_1$.
\vspace{16pt}

In summary, from the study of the movement of the random walker
described in this
section one can identify different scales.
They are related to those
excitations of $\H$
that describe the movement in question.
One finds that the lower band of excitations (slow modes) is related to 
the path from F to U and the higher band (fast modes) to the
movement around the favored corner F. 
This is confirmed by the results of appendix B
for the model with absorbing configurations.
For $h_c$, we find a scaling function
involving only the system size $L$ as a scale,
and depending on $tL^{-1}$.
This scaling is analogous to
the case of conformal invariance
in critical equilibrium systems.

We want to point out that, of course, the described scaling properties can
also be found by investigating order parameters different from $\M$.
%%%%%%%%%%%%%%%%%%%%%%%%%%%%%%%%%%%%%%%%%%%%%%%%%%%%%%%%%%%%%%%%%%%%%%%%%%%%%%%%%%%%
\section{Conclusions \label{sectioncon}}

In this paper, we studied the spectrum 
of the time evolution operator
and other
time-dependent properties of the two-dimensional 
non-local random walker (\ref{toy1}).
This model is a simple example of a class of non-equilibrium models
which exhibit spontaneous breaking of a {\boldmath $Z_2$} symmetry.
%Actually, the model is the zero-temperature limit
%of the three-state model of Ref.\cite{orig,us} and we used it to
%study non-stationary as well as stationary properties analogous to 
%the properties of the low temperature 
%phase of the three-state model.

The reduced size of the configuration space of the single particle model
of this paper allowed us to determine the spectrum numerically with enough precision
to extract the gaps (time scales) and the large $L$ limit of the low excitations.
For many-particle models a numerical investigation of the spectrum
is much more difficult and has not been done.
The investigation of the spectrum of the RWT model
led us to the following discoveries.

Breaking the symmetry explicitly with a small $h>0$, one is left with two
phases, a meta-stable and a stable one, in accordance with the fact
that the time evolution operator of the model has two ground states 
in the large $L$ limit. Above the two ground states, one has
two bands of excitations corresponding to two time scales.

At the point $h_c=\frac{a}{1-a}$, the meta-stable phase becomes unstable and, above that point,
one is left with only one ground state and correspondingly one (stable) phase.
This spinodal point has very interesting properties.
Fixing $h=h_c$, not only does one of the time scales vanish, but the
corresponding excitations also acquire the form $E_i=a\,k_i/L$.
Hence, it is universal in the sense that the
constants $k_i$ do not depend on the remaining free parameter of the model. 

At $h=h_c$,
the dynamical critical exponent is $z=1$, and the time scale vanishes with an exponent $x=2$.
Looking for ``conformal towers'', we did not find any regularities 
in the real part of the spectrum, but could show that the imaginary parts of the $k_i$ have
equidistant seperations.
We believe that the $1/L$ dependence on the system size is a more general phenomenon
($z=1$ for the first excitation had already been found for the three states model in
\cite{us})
and expect more structure to be found for the spectrum of many-particle-models at their
spinodal points.
The missing ``conformal towers'' for the RWT model may be a result of
the reduced configuration space.
Whether such a spectrum has further deep implications, as would be the case 
for equilibrium problems where one has conformal invariance, remains uncertain.

Besides the numerical diagonalization, we performed simulations of the
model to identify the physical relevance of the time scales. The lower gap
is linked with the movement from unfavored to favored phase
(slow modes), the higher gap
with the movement within the favored phase (fast modes). 
We could identify different scaling
regimes for the movement of the random walker starting at specific sites.
The scaling functions reflect the time scales found for the hamiltonian.
At the spinodal point, the appropriate scaling
variable for the movement starting from the unfavored phase
is $t/L$ and the corresponding scaling function is universal.

We also measured the first passage time for the random walker between stable and meta-stable phase.
We found that, below the critical point, both these flip times increase exponentially
with $L$. The flip times can be computed from the excitations of
a modified model with absorbing configurations (appendix B).
At the critical point, the first 
passage time from the unfavored to the favored corner of the triangle
has a non-trivial, universal linear dependence on $L$, 
consistent with the dynamical critical exponent $z=1$. 
The proportionality constant is different, by a factor of two, from its value above the spinodal
point where the flip time also increases linearly with $L$.
(There it can be calculated by a simple argument.)

As with the three-state model,
we have also investigated the free energy functional as
a function of the order parameter $d$ (\ref{orpa}).
Surprisingly for a non-local model without detailed balance
(but analogous to the three-state model),
we observed a link
between steady state properties (the free energy functional)
and non-stationary dynamics (flip times).

Those of the described results that relate to the region $h<h_c$
can also be found in a simplified, one-dimensional random walker model
(RWRA model, see appendix~C).
For this model one can calculate the free energy functional,
first excitations and flip times analytically.
The further reduction of the configuration space
leads to the different critical exponent $x=1$
instead of $x=2$.

%We cannot be sure how general are our results on the spectrum of the random walker model.
%But we believe that new and interesting critical and universal properties can be found 
%also by looking at the spectrum of many-particle models associated with symmetry breaking.
%To do this, though, one would have to implement new methods for studying the spectrum
%analytically or numerically (e.g.
%the density matrix renormalization group method\cite{DMRG}).

\ack
We would like to thank U.~Bilstein, C.~Godr\`{e}che, D. Mukamel
and especially V.~Rittenberg for helpful discussions and R.~Behrend
for reading the manuscript.
%%%%%%%%%%%%%%%%%%%%%%%%%%%%%%%%%%%%%%%%%%%%%%%%%%%%%%%%%%%%%%%%%%%%%%%%%%%%%%%%%%%%
\appendix
%%%%%%%%%%%%%%%%%%%%%%%%%%%%%%%%%%%%%%%%%%%%%%%%%%%%%%%%%%%%%%%%%%%%%%%%%%%%%%%%%%%%
\section{Movement of the random walker for $h>h_c$ 
\label{appendixvelocity}}

In the interior of the triangle ${\cal T}_L$, the random walker 
defined by (\ref{toy1}) has an average
velocity of
\begin{equation}
{v}=\left(\begin{array}{c}
-h-a(1+h)\\
h-a(1-h)
\end{array}
\right)
\end{equation}
independent of the site $(j,k)$.
Here, the first coordinate is along the $j$-axis 
(number of $-$ particles)
and the second along the $k$-axis ($+$ particles).
The projection onto the direction of the order parameter $\M$ is given by
\begin{equation}
\label{vorderparameter}
{v}_{\M}=-2 h(1-a)
\end{equation}

Repeating the argument of Ref.\cite{toy}, one sees that for small $h$
both components of ${v}$ are negative.
The walker near either of the corners $(L-1,1)$ or $(1,L-1)$
drifts towards the $j$- or $k$-axis and subsequently 
jumps back to the respective corner.
At $h_c=\frac{a}{1-a}$, the velocity is antiparallel to the 
$j$-axis.
For $h>h_c$, the second component of $v$ is positive,
rendering the unfavored corner around $(L-1,1)$ unstable,
because the random walker moves away from the $j$-axis
and
the non-local jumps along
this axis become irrelevant in leading
order. 
We have checked this with Monte Carlo simulations,
(see also section~\ref{sectionmove}). 

Neglecting these non-local jumps, from the mean path given by ${v}$
one can easily calculate the time taken
for a walker starting at
$\mbox{\U}=(L-1,1)$ to hit the $k$-axis and
to reach the favored corner.
One finds the flip time (\ref{fliphigh}).
In the same way,
(\ref{vorderparameter}) explains the leading term of (\ref{beginhigh})
and (\ref{eta}).

%%%%%%%%%%%%%%%%%%%%%%%%%%%%%%%%%%%%%%%%%%%%%%%%%%%%%%%%%%%%%%%%%%%%%%%%%%%%%%%%%%%%
\section{The RWT model with absorbing configurations \label{sectionabs}}

In this appendix, we investigate further the relation
of flip times and spectra.
We focus on the case where the time that the walker remains in a phase
increases exponentially with the system size.

The flip times $T_{\rm{long}}$ (\ref{flipmasslong}) and 
$T_{\rm{short}}$ (\ref{flipmassshort}) are defined as first passage times.
It is known that
the corresponding time scales are related to dynamical properties, i.e.
excitations, of a modified model with an absorbing state \cite{KaGa}. 
An absorbing state is a configuration
the system cannot escape from.
Here we have to choose 
U (for $T_{\rm long}$) or F (for $T_{\rm short}$) 
as the absorbing configuration.

We shall first present the general argument (see \cite{KaGa}).
Consider a reaction diffusion
system with an absorbing configuration $\{\beta_{\rm abs}\}$ and
described by a master equation (\ref{maeq}) with a
hamiltonian $H_{\rm abs}$.
The system may start in a configuration $\{\beta_0\}$ at time $t=0$:
\begin{equation}
|P(t=0)\!>_{\{\beta\}}=\left\{\begin{array}{ll}
                               1\mbox{ for }\{\beta\}=\{\beta_0\}\\
                               0\mbox{ for }\{\beta\}\neq\{\beta_0\}
                               \end{array}
                       \right.
\end{equation}
where the subscript of a vector denotes the component ${\{\beta\}}$.
%of vector ${\bf w}$.
Let $E_{\lambda}$ and $|P_{\lambda}\!>$ be
eigenvalues and eigenvectors of the hamiltonian
($E_0=0$ being the ground state energy). 
Note that the ground state $\Ps$ is given by 
\begin{equation}
|P_0\!>_{\{\beta\}}=\left\{\begin{array}{ll}
                               1\mbox{ for }\{\beta\}=\{\beta_{\rm abs}\}\\
                               0\mbox{ for }\{\beta\}\neq\{\beta_{\rm abs}\}
                               \end{array}
                       \right.
\end{equation}
If $H_{\rm abs}$ is diagonalizable,
there exist
$a_{\lambda}\in {\bf C}$ such that
\begin{equation}
|P(t=0)\!>=\sum_{\lambda}\,a_{\lambda}\;|P_{\lambda}\!>\;\;\;.
\end{equation}

Obviously, the probability $P_{\rm in}$
that the system has not reached its absorbing state is given by
\begin{eqnarray}
P_{\rm in}(t)&=1-P(\{\beta_{\rm abs}\};t)\nonumber\\
           &=1-{\left(\exp(-H_{\rm abs}\,t)\,
           |P(t=0)\!>\right)}_{\{\beta_{\rm abs}\}}\\
	   &=1-\sum_{\lambda} a_{\lambda}\,
             \exp(-E_{\lambda}\,t)\,|P_{\lambda}\!>_{\{\beta_{\rm abs}\}}\nonumber\;.
\end{eqnarray}
Denote as ${p}(t)\d t$ the probability that the system reaches $\{\beta_{\rm abs}\}$ 
during the time interval $[t,t+\d t]$. Then
\begin{equation}
p(t)\d t=P_{\rm in}(t)-P_{\rm in}(t+\d t)\;\;\;.
\end{equation}
Thus, ${p}(t)$
is given by
\begin{equation}
{p}(t)=-\frac{\mbox{d}P_{\rm in}(t)}{\mbox{d}t}=-
\sum_{\lambda}\,a_{\lambda}
E_{\lambda}\,\exp(-E_{\lambda}\,t)\;|P_{\lambda}\!>_{\{\beta_{\rm abs}\}}\; ,
\end{equation}
and its mean value, the flip time, is
\begin{equation}
T=-\sum_{\lambda\neq 0}\frac{a_{\lambda}
\;|P_{\lambda}\!>_{\{\beta_{\rm abs}\}}}{E_{\lambda}} \;.
\end{equation}
From the last equation, it follows,
under the conditions
\begin{eqnarray}
\fl
E_1\propto\exp(-m\,L)\, ,\qquad 
\lim_{L\rightarrow\infty} E_i>0 \mbox{ for }i\neq 0,1\, ,
\qquad \exists y>0: \lim_{L\rightarrow\infty}L^y a_1\neq 0\; , \nonumber
\end{eqnarray}
that in leading order
\begin{eqnarray}
\label{twophases}
T(L)\propto\exp(m\,L)\;.
\end{eqnarray}
The result is similar for any finite number of excited states 
$E_i\propto\exp(-m_i\,L)$.
In that case, the scale of the flip time is given by the largest $m_i$, with
$\lim_{L\rightarrow\infty}L^y a_i\neq 0$ for some~$y$.

On the other hand, if none of the excited levels approaches zero
for $L\rightarrow\infty$,
one has
\begin{equation}
\label{onephase}
T(L)\leq {\cal O}\left({\mbox{dim}(L)}\right)
\end{equation}
independent of $\{\beta_0\}$ ($\mbox{dim}(L)$ is the dimension of the
configuration space for system size $L$).
This implies that there is only one phase in the large $L$ limit.

We now turn to specific results for the RWT model.
We have diagonalized the modified hamiltonian numerically.
From the spectra
we find the length scales $\mu_{\rm long}$ and $\mu_{\rm short}$
of the flip times:
\begin{itemize}
\vspace{6pt}
\item
With the favored corner \F\ made absorbing, the scale $\omega$ (\ref{lev1})
remains unchanged 
as does the gap of the lower band of excitations $m_1$.
This explains the relation 
$\mu_{\rm short}=\omega$ (\ref{flipeq}) which is a special case of (\ref{twophases}).
Similar relations are found analytically for the simplified model 
discussed in appendix C.

The lower gap vanishes at the critical point $h_c$
and the $1/L$ dependence of the levels is the same as for the non-absorbing RWT model
(table \ref{tabcritlev}).
The upper gap changes to a new value $m_3>m_2$. 

These results confirm the interpretation that the 
second ground state (which exists for $h<h_c$, $L\rightarrow\infty$)
and the lower band of excitations are
linked with the meta-stable phase and 
with the movement from the unfavored
to the favored corner. The corresponding scales of the spectrum
remain unchanged,
while the movement within the favored phase, and correspondingly
the upper band of excitations, is altered
by the absorbing configuration \F .
\vspace{6pt}
\item
In the case of an absorbing configuration \U\ in the (originally)
unfavored phase,
there is more change to the spectrum.
There is a first excited state $E_1\propto \exp (-{\omega}'\, L)$ with 
$\omega'\neq\omega$ for any $h$,
connected with two steady states for $L\rightarrow\infty$.
This reflects the fact that, with absorption in the corner U,
both corners U and F constitute a phase for any $h$.
From (\ref{twophases}) one has
\begin{equation}
\mu_{\rm long}={\omega}' \;\;\mbox{    for any  }h,
\end{equation}
analogous to (\ref{flipeq}).
This relation is consistent with the numerical data to a
precision of about $15\%$.
 
Further, the scale ${\omega}'$ is related to $m_2$ of the RWT model without an absorbing
configuration.
For $a=0.25$ one has
\begin{equation}
{\omega}'=2.5(2)\,m_2\;\;\;.
\end{equation}

As a further change to the spectrum,
the lower band of excitations is shifted to
a new band of excitations with a gap $m_3>m_2$, while
the gap $m_2$ remains
unchanged.

The new upper band of excitations is related to
movement within the former unfavored corner.
Note that $m_3(h)$ takes the same
values with either the favored or the
unfavored corner made absorbing. The corresponding eigenstates
describe the (rapid) movement near an absorbing corner.
\end{itemize}
%
%%%%%%%%%%%%%%%%%%%%%%%%%%%%%%%%%%%%%%%%%%%%%%%%%%%%%%%%%%%%%%%%%%%%%%%%%%%%%%%%%%%%
\section{Analytical results form the RWRA model \label{appsupertoy}}
%
%
%In this appendix we discuss a simplification of the RWT model.
%We take 
%
%\begin{equation}
%b=0
%\end{equation}
%
%and keep only the points $(j,k)\in {\cal T}_L$
%with $j=1$ or $k=1$.
%That is, we consider only movement along the
%$j$- and $k$-axes, where macroscopic jumps occur.
%We refer to this model as the RWRA model (random
%walker on a right angle).
%In an infinitesimal time interval the following processes
%can occur with a probability $p$:
%
%\begin{eqnarray}
%\label{sutoy}
%(j,1)\rightarrow (j-1,1)\qquad &\mbox{if $j>1$ with }p=a(1+h)\d t
%\nonumber\\
%(j,1)\rightarrow (L-1,1)\qquad &\mbox{with }p=c(1-h)\d t
%\nonumber\\
%(1,k)\rightarrow (1,k-1)\qquad &\mbox{if $k>1$ with }p=a(1-h)\d t
%\nonumber\\
%(1,k)\rightarrow (1,L-1)\qquad &\mbox{with }p=c(1+h)\d t
%\end{eqnarray}
%
%The RWRA model is essentially a one-dimensional (albeit non-local)
%random 
%walker model and one is able to calculate exactly
%the flip times from stable to meta-stable corner
%and vice versa,
%the exponential decay of the first excitation with $L$
%and
%the \FEF\ as a function of the
%order parameter $\M$.
%Again we limit our investigation to the case $c=\frac12$
%(this fixes the unit of time) 
%and $a\equiv a+b=\frac12$. 
%Taking $a$ different from $\frac12$ does not change 
%the physics of the RWRA model.

%For the spinodal point of the RWT model one has
%$\lim_{a\rightarrow 1/2}h_c=1$.
%Therefore, the RWRA model, describing part of the RWT
%model for $a\rightarrow 1/2$, mimics the physics of the
%RWT model below the critical point
%and one expects
%a stable and a meta-stable phase 
%for any $h$.

In this appendix, we investigate the model defined by Eq.(\ref{sutoy}). It is expected to
give the physics of the RWT model below the spinodal point $h_c$.
We
shall indeed see that for
the RWRA model there are first passage
times from one corner to the other which increase exponentially with $L$ as
for the RWT model.
We calculate
their exact expressions (Eqs.(\ref{stflipshort}) and (\ref{stfliplong}))
and find the two scales with which these flip times diverge.

We are also able to compute the first excitation (Eq.(\ref{Ei})), which becomes
a second ground state in the large $L$ limit,
and the first excitation for the RWRA model with either 
of the corners
made absorbing (Eqs.(\ref{Eii}) and (\ref{Eiii})). 
The first excitations are proportional
to $\exp(-\omega L)$. In the case of no absorption
or absorption at the favored corner, $\omega$ is equal
to the scale of the short flip time (\ref{stfliplong});
and, in the case
of the unfavored corner made absorbing, $\omega$ is equal 
to the scale of the long flip time (\ref{stfliplong}).
All these results are analogous to those for the RWT model, but
here they are obtained analytically.

We have also computed the free energy functional Eq.(\ref{stfreee}),
whose graph is very similar to that for the RWT model's.
As in the RWT model, the height differences 
$f_{\rm max}-f_{\rm min} $ give the scales with which the flip times 
(\ref{stflipshort}--\ref{stfliplong})
diverge.

\subsection{Flip times}
As for the RWT model, for $h=0$ the random walker 
stays most of the time near the
points F=$(1,L-1)$ and U=$(L-1,1)$.
For $h>0$, the point F is favored and the stable 
phase is concentrated
around this site.

To calculate the flip time from the meta-stable to the 
stable phase, we define $D_{j,k}$ as the first passage time
from $(j,k)$ to $(1,L-1)$.
%The flip time from the stable to the meta-stable phase 
%is then $T_{\rm short}=D_{L-1,1}$.

Concentrating on the 
first step of the random walker (in an infinitesimal
time interval ${\rm d}t$), we find 
\begin{eqnarray}
D_{1,k}=&
(1-\d t )(D_{1,k} + \d t)
\nonumber\\
&+\hm \d t (D_{1,k-1} + \d t)
\nonumber\\
&+\hp \d t (D_{1,L-1} + \d t). 
\nonumber
\end{eqnarray}
for $j=1, k=2,\dots,L-2$ (cf. \cite{toy}).
This yields
\begin{equation}
D_{1,k}=1 + \hm D_{1,k-1} +
          \hp D_{1,L-1}.
\end{equation}
Similarly, the other conditions for
the $D_{j,k}$ read
\begin{eqnarray}
D_{j,1}=1 + \hp D_{j-1,1} +\hm D_{L-1,1}  
\qquad\qquad j=2,\dots,L-2
\label{recurj}
\\
D_{1,1}=1 + \hp D_{1,L-1} + \hm D_{L-1,1}
\\
\hm D_{1,L-1} = 1+ \hm D_{1,L-2}
\\
\hp D_{L-1,1} = 1+ \hp D_{L-2,1}\; .
\label{bc}
\end{eqnarray}
From the definition of $D_{j,k}$, the boundary conditions are given by
\begin{eqnarray}
D_{1,L-1}=0
\nonumber
\\
\label{boundary2}
D_{L-1,1}=\Ts \;\;.
\end{eqnarray}
The recurrence relation (\ref{recurj}) can be solved as
\begin{equation}
D_{j,1}= D_{1,1} \frac{2}{1-h} 
%\left( (1-(\hp)^2)(\hp)^{j-2}+1 \right).
\left( \left(\hp\right)^{j-2} - \left(\hp\right)^j+1 \right)\; ,
\end{equation}
and the condition (\ref{bc}) gives us the flip time 
\begin{equation}
\label{stflipshort}
T_{\rm short}=
\left(\frac{2}{1-h}\right)^2
\left(\frac{2}{1+h}\right)^{L-2}
-\frac{2}{1-h} \; .
\end{equation}
Replacing $h$ by $-h$, we obtain the flip time 
from favored to unfavored phase
\begin{equation}
\label{stfliplong}
T_{\rm{long}}=
\left(\frac{2}{1+h}\right)^2
\left(\frac{2}{1-h}\right)^{L-2}
-\frac{2}{1+h}.
\end{equation}
One sees that the flip times diverge exponentially with $L$.
The scales are $\mu_{\rm{short}}=\log(2/(1-h))$ and
$\mu_{\rm{long}}=\log(2/(1+h))$. This implies that there are
two ground states in the large $L$ limit.
Linear combinations
of these give the phases; a stable phase concentrated around F$=(1,L-1)$,
and a meta-stable around U=$(L-1,1)$.
%
%%%%%%%%%%%%%%%%%%%%%%%%%%%%%%%%%%%%%%%%%%%%%%%%%%%%%%%%%%%%%%%%%
\subsection{$L$ dependence of the first excitation}
With the notation $x_s=P(s,1)$, $y_s=P(1,s)$ and $x_1=y_1=P(1,1)$,
the master equation 
of the RWRA model reads
\begin{eqnarray}
\label{xxxxx}
\frac{\d}{\d t}x_s=-x_s+\hp x_{s+1}
\qquad\qquad\qquad\qquad &s=2,\dots,L-2
\\
\frac{\d}{\d t}y_s=-y_s+\hm y_{s+1}
                         &s=2,\dots,L-2
\label{marey}
\\
\frac{\d}{\d t}x_1=-x_1+\hp x_2 +\hm y_2
\\
\frac{\d}{\d t}x_{L-1}=-x_{L-1} 
+ \hm \sum_{i=1}^{L-1}x_i
\\
\frac{\d}{\d t}y_{L-1}=-y_{L-1} 
+ \hp \sum_{i=1}^{L-1}y_i
\label{mabcy} \; .
\end{eqnarray}
Setting $x_s=\exp(-Et)X_s$ and $y_s=\exp(-Et)Y_s$, 
%
%\begin{eqnarray}
%(1-E)X_s=\hp X_{s+1}
%\qquad\qquad\qquad\qquad &s=2,\dots,L-2
%\label{recx}
%\\
%(1-E)Y_s=\hm X_{s+1}
%&s=2,\dots,L-2
%\label{recy}
%\\
%(1-E)X_1=\hp X_2 +\hm Y_2
%\\
%(1-E)X_{L-1}=\hm \sum_{i=1}^{L-1}X_i
%\label{bcx}
%\\
%(1-E)Y_{L-1}=\hp \sum_{i=1}^{L-1}Y_i
%\label{bcy}
%\end{eqnarray}
%
%
%After solving the recurrence relations (\ref{recx}) and (\ref{recy}),
and solving the resulting recurrence relations,
%the boundary conditions (\ref{bcx}) and (\ref{bcy}) give the 
we find
the condition on an eigenvalue $E$,
\begin{eqnarray}
0=&E\Bigg(
E(1-E)^{2L-3}
+\frac{(1-h^2)^{L-1}}{4^{L-1}}
\nonumber\\
&\quad-(1-E)^{L-2} (1-h^2) \frac{(1+h)^{L-2} + (1-h)^{L-2}}{2^{L}}
\Bigg)
\nonumber\\
&\Bigg(
(1-E)^{2L-5}(\hp -E)(\hm -E)
\Bigg)^{-1}
\; .
\end{eqnarray}
There is one solution with $E_0=0$.
The corresponding eigenvector is
the steady state for finite $L$.
For $L\gg 1$, the first excited level behaves
as
\begin{equation}
E_1
\cong\frac14 (1-h^2)\left(\hp\right)^{L-2}
\propto \exp(\omega L) \; .
\label{Ei}
\end{equation}
Thus, 
for
$L\rightarrow\infty$,
one has two stationary states resulting in two phases.

The higher spectrum has a two band structure similar to that for the RWT model.
The corresponding gaps are $m_1=(1-h)/2$ and $m_2=(1+h)/2$.
Thus, one finds 
$\omega =m_1$ for $h\rightarrow 1$.
(For the RWT model one has proportionality of $\omega$ and $m_1$
for any $h$, see (\ref{gapsexp}).)
We point out that for the
RWRA model the gap vanishes with a critical exponent $x=1$,
while for the RWT model one finds $x=2$ (see Eq.(\ref{gapsexp})).
This is due to the reduction of the configuration space.

One can compute the large $L$ behavior of the flip times
from the $L$-dependence of the first
excited level of a modified model
(appendix B).
For the flip time $T_{\rm short}$,
one implements an  
absorbing site at the stable corner $(1,L-1)$,
replacing equations (\ref{marey}) and (\ref{mabcy})
by
\begin{eqnarray}
\frac{\d}{\d t}y_s=-y_s+\hm y_{s+1}
\qquad\qquad\qquad\qquad s=2,\dots,L-3
\\
\frac{\d}{\d t}y_{L-2}=-y_{L-2}
\\
\frac{\d}{\d t}y_{L-1}=\hp \sum^{L-2}_{i+1}y_i \; .
\end{eqnarray}
With this, one finds a first excited level
\begin{equation}
E_1^{(1,L-1) \rm  abs.}\cong\frac14 (1-h^2)\left(\hp\right)^{L-2}.
\label{Eii}
\end{equation}
With $(L-1,1)$ absorbing, instead of $(1,L-1)$, one analogously obtains
\begin{equation}
E_1^{(L-1,1) \rm  abs.}\cong\frac14 (1-h^2)\left(\hm\right)^{L-2}.
\label{Eiii}
\end{equation}
This way, with equations
(\ref{stflipshort}),(\ref{stfliplong})
and (\ref{twophases}), we find
\begin{eqnarray}
\Ts\propto \left(\frac{2}{1+h}\right)^{L-2}
\nonumber\\
\Tl\propto \left(\frac{2}{1-h}\right)^{L-2}
\nonumber \; ,
\end{eqnarray}
which is indeed consistent with the flip times (\ref{stflipshort}) and
(\ref{stfliplong}) computed directly above.
%%%%%%%%%%%%%%%%%%%%%%%%%%%%%%%%%%%%%%%%%%%%%%%%%%%%%%%%%%%%%%%%
\subsection{The stationary probability distribution}
To calculate the probability distribution of the steady state,
one has to solve equations (\ref{xxxxx})--(\ref{mabcy}) for 
$\frac{\d}{\d t}x_s=\frac{\d}{\d t}y_s=0$.
One finds 
\begin{eqnarray}
X_s=\frac14 N (1-h^2) \left(\frac{2}{1+h}\right)^s
\\
Y_s=\frac14 N (1-h^2) \left(\frac{2}{1-h}\right)^s
\qquad\qquad\qquad\qquad s=2,\dots,L-1
\\
X_1=Y_1=N \; ,
\end{eqnarray}
where the normalization $N$ is
\begin{equation}
N=\frac{(1-h^2)^{L-2}}{(2(1-h))^{L-2}+(2(1+h))^{L-2}-(1-h^2)^{L-2}}\; .
\end{equation}
The free energy, defined by (\ref{deffreeenergy}),
takes the form
\begin{equation}
\label{stfreee}
f(\M)=\left\{
\begin{array}{ll}
\log\left(\frac2{1-h}\right)
   -\M \log\left(\frac2{1+h}\right)
&{\rm for\ } \M\geq 0\\[2mm]
\log\left(\frac2{1-h}\right)
   +\M \log\left(\frac2{1-h}\right)
&{\rm for\ } \M\leq 0\; .\\
\end{array}
\right.
\end{equation}
The minima of the free energy functional are at $\M=-1$ and $\M=1$,
corresponding to the stable and metastable phase respectively.
In between, the free energy functional has a constant slope.

In conclusion, we have shown that the RWRA model presented in this appendix
captures most of the physics of the RWT model
below the critical point.
%%%%%%%%%%%%%%%%%%%%%%%%%%%%%%%%%%%%%%%%%%%%%%%%%%%%%%%%%%%%%%%%%%%%%%%%%%%%%%%%%%%%
\newpage
%%%%%%%%%%%%%%%%%%%%%%%%%%%%%%%%%%%%%%%%%%%
%     bibliography                        %
%%%%%%%%%%%%%%%%%%%%%%%%%%%%%%%%%%%%%%%%%%%
%
%\newpage

%
%%%%%%%%%%%%%%%%%%%%%%%%%%%%%%%%%%%%%%%%%%%%%%%%%%%%%%%%%%%
%
\end{document}